\documentclass[12pt]{article} 
\usepackage{amsmath}
\usepackage{amssymb}
\usepackage{amsfonts}
\usepackage{multirow}
\usepackage{natbib}
\usepackage{amsthm}
\usepackage{graphicx}
\usepackage{booktabs}
\usepackage[tight]{subfigure}
\usepackage{caption}
\usepackage{url}
\usepackage{mathrsfs}
\usepackage{xcolor}
\usepackage{mathtools}
\usepackage{graphicx}
\usepackage{setspace}
\usepackage[margin=1in]{geometry}
\usepackage{float}

\newcommand{\be}{\begin{equation}}
\newcommand{\ee}{\end{equation}}
\newcommand{\bea}{\begin{eqnarray}}
\newcommand{\eea}{\end{eqnarray}}
\newcommand{\mbf}[1]{\mathbf{#1}}
\newcommand{\mbs}[1]{\boldsymbol{#1}}

\newcommand{\bs}{\mbf{s}}

\newcommand{\bz}{\mbf{z}}

\newcommand{\bsigma}{\mbs{\sigma}}
\newcommand{\btau}{\mbs{\tau}}

\newcommand{\bmu}{\mbs{\mu}}
\newcommand{\bbeta}{\mbs{\beta}}

\newcommand{\balpha}{{\mbs{\alpha}}}
\newcommand{\bepsilon}{{\mbs{\epsilon}}}

\newcommand{\btheta}{{\mbs{\theta}}}
\newcommand{\bseta}{{\mbs{\eta}}}

\newcommand{\ben}{\begin{equation*}}
\newcommand{\een}{\end{equation*}}
\newcommand{\bean}{\begin{eqnarray*}}
\newcommand{\eean}{\end{eqnarray*}}
\newcommand{\bsm}{\begin{smallmatrix}}
\newcommand{\esm}{\end{smallmatrix}}
\newcommand{\bmat}{\begin{matrix}}
\newcommand{\emat}{\end{matrix}}

\newcommand{\given}{\,|\,}

\newcommand{\bell}{\mbs{\ell}}

\begin{document}

    \begin{center}
        \vspace*{1cm}
        \large
	    \textbf{Quantifying and correcting geolocation error in spaceborne LiDAR forest canopy observations using high spatial accuracy ALS: A Bayesian model approach}\\
         \normalsize
           \vspace{5mm}
	    Elliot S. Shannon\textsuperscript{1, 2}, Andrew O. Finley\textsuperscript{1,2}, Daniel J. Hayes\textsuperscript{3},\\Sylvia N. Noralez\textsuperscript{3}, Aaron R. Weiskittel\textsuperscript{3}, Bruce D. Cook\textsuperscript{4}, Chad Babcock\textsuperscript{5}
	   \\
         \vspace{5mm}
    \end{center}
    {\small 
    \begin{enumerate}
        \item Department of Forestry, Michigan State University, East Lansing, MI, USA.
        \item Department of Statistics and Probability, Michigan State University, East Lansing, MI, USA.
        \item School of Forest Resources, University of Maine, Orono, ME, USA. 
        \item Biospheric Sciences Laboratory, NASA Goddard Space Flight Center, Greenbelt, MD, USA.
        \item Department of Forest Resources, University of Minnesota, Saint Paul, MN, USA.
        \end{enumerate}
        }
        \noindent \textbf{Corresponding Author}: Elliot S. Shannon, email: shann125@msu.edu.\\

\section*{Abstract}


Geolocation error in spaceborne sampling light detection and ranging (LiDAR) measurements of forest structure can compromise forest attribute estimates and degrade integration with georeferenced field measurements or other remotely sensed data. Data integration is especially problematic when geolocation error is not well quantified.  We propose a general model that uses airborne laser scanning (ALS) data to quantify and correct geolocation error in spaceborne sampling LiDAR. To illustrate the model, LiDAR data from NASA Goddard's LiDAR Hyperspectral \& Thermal Imager (G-LiHT) was used with a subset of LiDAR data from NASA's Global Ecosystem Dynamics Investigation (GEDI). The model accommodates multiple canopy height metrics derived from a simulated GEDI footprint kernel using spatially coincident G-LiHT, and incorporates both additive and multiplicative mapping between the canopy height metrics generated from both datasets. A Bayesian implementation provides probabilistic uncertainty quantification in both parameter and geolocation error estimates. Results show a systematic geolocation error of 9.62 m in the southwest direction. In addition, estimated geolocation errors within GEDI footprints were highly variable, with results showing a $\sim$0.45 probability the true footprint center is within 20 m. Estimating and correcting geolocation error via the model outlined here can help inform subsequent efforts to integrate spaceborne LiDAR data, like GEDI, with other georeferenced data.

\section{Introduction}

Models of forest attributes such as aboveground biomass (AGB) are becoming increasingly important tools for monitoring forest carbon storage and sequestration. To this end, accurate measurements of forest structure represent an invaluable resource for both research and management of forest ecosystems. Traditionally, sampling techniques including field surveys and airborne laser scanning (ALS) campaigns have provided detailed measurements of 3D forest structure, often with relatively small geolocation error. However, these measurements are typically limited in their temporal and spatial extent due to high acquisition cost. Now, a growing number of spaceborne light detection and ranging (LiDAR) campaigns provide forest structure measurements at nearly global extent, allowing new opportunities for multi-sensor data fusion in AGB models. However, the geolocation error of measurements from these spaceborne sensors can be large and not well quantified, leading to increased uncertainty in model estimates. Given this trade-off, models of forest structure can benefit from incorporating both airborne and spaceborne LiDAR data, with careful consideration given to their integration \citep{silva_et_al_2021}.

 Global spaceborne LiDAR systems are a revolutionary technology used to systematically measure the Earth's surface as well as vegetation structure. A recent example is NASA's Global Ecosystem Dynamics Investigation (GEDI) LiDAR mission, which provides measurements of forest structure along the orbital path of the International Space Station (ISS). Like other spaceborne LiDAR systems, GEDI takes samples at small, evenly spaced footprints (25 m) along ground tracks using high resolution waveform LiDAR, which provides 3D measurements of the Earth's surface including ground elevation, forest canopy height, and forest canopy cover \citep{dubayah_et_al_2020}. Spaceborne LiDAR data collected in this way have been used for many years in efforts to map and model forest ecosystems, and since its launch in December of 2018, GEDI data have been extensively used for applications such as carbon monitoring and forest AGB estimation \citep{saarela_et_al_2018, silva_et_al_2021, duncanson_et_al_2022}, often informed using ancillary remote sensing (ALS) and field data. Specifically, relative height (RH) metrics, which represent the vertical distribution of forest canopy heights above ground, are calculated at a number of percentiles across the canopy elevation profile, and serve as key explanatory covariates in AGB models. Spaceborne LiDAR systems benefit greatly from their ability to capture data at a nearly global extent, and by repeatedly sampling transects along the ISS orbit, GEDI's extensive spatial coverage provides great opportunity for modeling applications over large scales. However, due to factors including positioning uncertainty of the sensor, the orientation of the sensor to the satellite, and image processing errors, geolocation uncertainty of spaceborne LiDAR data may be increased \citep{roy_et_al_2021}. For GEDI, the geolocation error for the Version 2 data is estimated to be $\sim$10 m \citep{dubayah_et_al_2020}. 

 In contrast to spaceborne LiDAR systems, ALS uses aircraft-mounted instruments to capture fine-resolution LiDAR data in specific focal areas. For example, Goddard's LiDAR, Hyperspectral, and Thermal Imager (G-LiHT) is a multi-sensor, airborne remote sensing system that maps forest vertical structure including RH metrics \citep{cook_et_al_2013}. As an airborne instrument, the temporal and geographic extent of ALS systems such as G-LiHT are limited compared to spaceborne sensors such as GEDI. However, G-LiHT’s airborne LiDAR scanner can acquire and discretize a higher sample density ($\sim$12 pulses m$^2$) of small footprint laser pulses ($\sim$10 cm diameter) to create a continuous, canopy height surface layer model at 1 m horizontal spatial resolution. Moreover, individual discrete return measurements used to compute vertical canopy heights and forest RH metrics for ALS systems benefit from very little positional error between repeat acquisitions ($<$10 cm for G-LiHT). 

Characterizing and correcting spaceborne LiDAR geolocation error is a crucial step when integrating these data with other sensor data and field measurements for applications such as AGB estimation and mapping. Persistent geolocation errors may undermine forest canopy height retrievals in areas of complex or heterogeneous forest structure \citep{frazer_et_al_2011, milenkovic_et_al_2017, roy_et_al_2021} and thus hinder identification of coincident measurements with field plot and ALS datasets \citep{duncanson_et_al_2022}. To this end, previous studies have implemented a variety of methods to characterize or even correct the geolocation uncertainty of spaceborne LiDAR data, including GEDI, and improve estimates of terrain and canopy heights, which are key covariates for forest AGB estimation \citep{liu_et_al_2021, quiros_et_al_2021, roy_et_al_2021, wang_et_al_2021}. In these studies, coincident ALS data with smaller geolocation errors are compared with spaceborne LiDAR measurements to assess disagreements between reported spaceborne RH metrics and observed ALS data, often via simulation to allow direct comparison. For example, in a waveform interpretation study involving GEDI, \cite{lang_et_al_2022} first implemented a geolocation correction step to optimally correlate simulated GEDI waveforms with ALS data. Here, Version 1 GEDI data were corrected in blocks, with an initial estimated mean horizontal positioning error of $\sim$20 m, which was reduced to only a few meters on average using the method outlined in \cite{hancock_et_al_2019}. Similarly, \cite{wang_et_al_2021} compared the root mean squared error (RMSE) values of GEDI performance with and without colocation calibration, and observed increased uncertainty in estimated top of canopy RH values for uncalibrated footprints. Therefore, quantification or correction of geolocation error in spaceborne LiDAR is crucial for propagating measurement uncertainty and improving model estimates of forest attributes such as AGB, especially when combined with other data sources such as ALS \citep{wang_et_al_2021}.

 New methods have been developed in an attempt to better characterize and even correct geolocation error associated with spaceborne LiDAR data. In a previous study, \cite{liu_et_al_2021} used discrete return ALS data to simulate GEDI waveforms over a range of potential horizontal offsets, and quantified GEDI's geolocation error using the offset with the highest correlation between waveforms. This waveform matching approach resulted in slight improvements in both RMSE and mean absolute error (MAE) values for GEDI terrain height estimates. Similarly, \cite{quiros_et_al_2021} assessed the positional error of reported GEDI footprint locations using ALS data by considering 16 positional adjustments at distances of 5 or 10 m in the along and across track directions, as well as at intermediate angular positions. Here, the position resulting in the lowest RMSE was considered optimal, and the authors observed a consistent improvement in RMSE of estimated ground elevations for positional adjustments of $270^{\circ}$ within 10 m. These studies highlight the dual sources of geolocation error associated with spaceborne LiDAR measurements such as GEDI, which might result from both biases in the instrument itself, along with random noise at the level of individual footprints. Therefore, geolocation error assessments should aim to identify systematic shifts in footprint-level measurements that improve agreement with ALS measurements, while incorporating flexibility to capture location specific effects. 
 
The aforementioned approaches estimate geolocation error by optimizing a simple objective function (e.g., minimizing RMSE) that compares a sensor recorded metric to those derived using a higher spatial accuracy data source (e.g., ALS) at, and around, the sensor recorded locations. While effective at delivering a systematic geolocation error point estimate (e.g., distance and direction), these methods are not couched in a probabilistic framework and thus offer little in the way of uncertainty quantification for geolocation error and parameters used to map sensor metrics to those derived from higher spatial accuracy data sources, cannot easily partition location specific and systematic (i.e., across locations) geolocation error, and are not easily extendable to identify non-stationary geolocation error within a unified framework. 

 Other studies have investigated the impacts of spaceborne LiDAR geolocation error on measurements of forest canopy height and AGB estimates. \cite{frazer_et_al_2011} investigated the effects of co-registration error between simulated LiDAR and sample plot data on the accuracy and uncertainty of AGB estimates and observed a negative effect of co-registration error on estimation accuracy of total forest AGB. Additionally, \cite{roy_et_al_2021} used a Monte Carlo simulation technique to randomly generate positional errors of GEDI footprints from ALS data in the Congo. They found the impact of geolocation error on GEDI canopy height retrieval was dependent on local spatial variation in forest canopy height. In this way, spaceborne canopy height measurements of complex and fragmented forests may be more greatly impacted by geolocation error than those of more homogeneous forest types \citep{frazer_et_al_2011, milenkovic_et_al_2017}. As forests become more heterogeneous following increased disturbances and diverse management practices, statistical methods are urgently needed to assess the effect of geolocation error and propagate its associated uncertainty to estimates of forest attributes derived from spaceborne LiDAR data.   

 In this paper we propose and illustrate a Bayesian model that uses high spatial accuracy ALS data to quantify and correct geolocation error in forest canopy measurements from larger footprint spaceborne sampling LiDAR. Using this model, we aim to learn about both location specific and systematic (i.e., across locations) geolocation error using spatially-coincident ALS data. The proposed model: 1) uses a sampling LiDAR footprint kernel of known functional form with parameters that can be fixed or estimated; 2) considers multiple canopy metrics at each footprint location; 3) estimates additive and multiplicative mapping between canopy metrics generated from the two LiDAR data sources; 4) yields key summaries of geolocation error within and across locations where both LiDAR data sources are observed; 5) provides probabilistic uncertainty quantification in parameter and geolocation error estimates; and 6) extends in a straightforward way to accommodate geolocation error that might be non-stationary in space and/or time (e.g., settings where geolocation error might vary based on acquisition location or time).
 
While the proposed modeling framework is general and might be applied in a variety of settings, we motivate the method's exposition using the GEDI and G-LiHT data described in Section \ref{sec:data}. The proposed model is presented in Section~\ref{sec:models}. Here, too, we define a submodel and criterion to assess model fit and describe how we characterize geolocation error. Analysis results are presented in Section~\ref{sec:results}, which is followed by a discussion of results in Section~\ref{sec:discussion}. Some model extensions are described in Section~\ref{sec:conclusion}, along with concluding remarks and next steps.

\section{Data}\label{sec:data}

\subsection{Study area}

\begin{figure}[!ht]
     \centering
     \includegraphics[width=12cm]{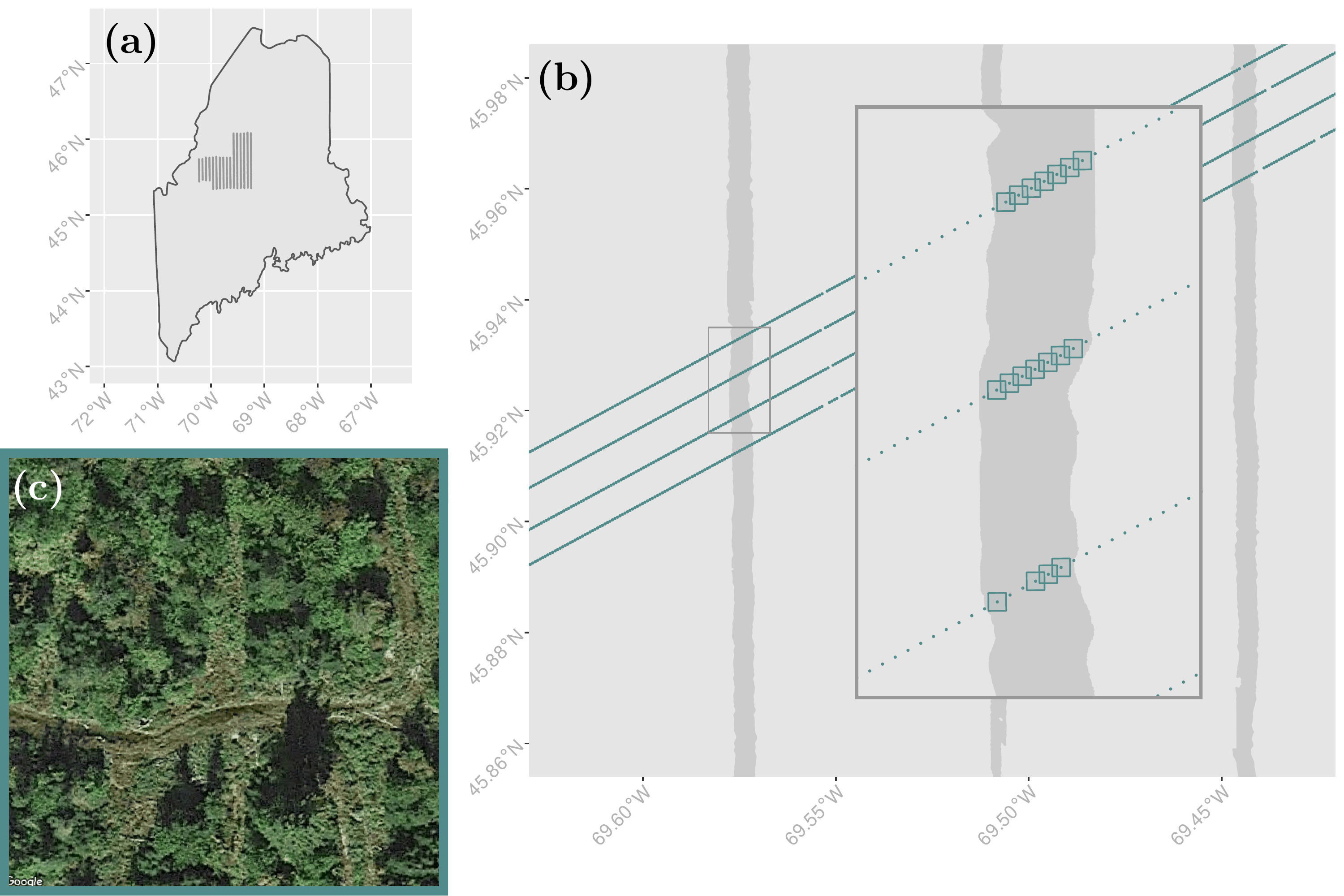}
     \caption{(a) Regional view of the Maine CMS study area, with G-LiHT airborne LiDAR acquisition tracks oriented in the north-south direction. (b) Sample of unfiltered spaceborne GEDI LiDAR data (green) captured over the G-LiHT acquisition tracks, with 70 m square buffers generated for spatially coincident, high-quality filtered GEDI observations. (c) Satellite imagery for a single, high-quality GEDI observation with 70 m square buffer.}
     \label{fig:study_area}    
 \end{figure}

The NASA Carbon Monitoring System (CMS) Maine study site is located in Northern Maine, USA (Figure~\ref{fig:study_area}a). As a unit of the North American Forest Dynamics Project \citep{goward_et_al_2008}, the Maine CMS area has been closely monitored via airborne and field forest inventories for nearly a decade \citep{deo_et_al_2017}. Most recently, ALS data coincident with GEDI spaceborne measurements were captured over the Maine CMS area (Figure~\ref{fig:study_area}b). The area is composed of a mix of both hardwood and softwood tree species and has a complex structure reflecting the unique disturbance history of varied logging and management strategies. These mixed harvest patterns have created a diverse forest canopy structure across the study site, with both fine and coarse scale forest canopy gaps present (Figure~\ref{fig:study_area}c). This fine-scale heterogeneity can exacerbate any potential disagreement between spaceborne and airborne canopy height measurements caused by geolocation error \citep{milenkovic_et_al_2017}.

\subsection{GEDI data}\label{sec:gedi}

Version 2 Global Footprint Level GEDI L2A Elevation and Height Metrics data were downloaded from the USGS Land Processes Distributed Active Archive Center (LP DAAC) Data Pool using a spatial query through NASA's Common Metadata Repository (CMR) (see Supplementary Material for detailed data download and processing steps). The GEDI L2A dataset represents a spaceborne LiDAR campaign to capture full waveform LiDAR data globally with a footprint diameter of $\sim$25 m and estimated geolocation error of $\sim$10 m \citep{dubayah_et_al_2020}. The GEDI data are reported as a collection of RH metrics at 100 percentiles, along with quality flags, and unique identifiers for each footprint. Given the estimated $\sim$10 m geolocation error of the Version 2 GEDI data, along with the $\sim$12.5 m radius of the GEDI footprint, a square focal area of side-length 45 m was centered on each reported GEDI footprint center. This focal area was taken as the search domain to capture the full extent of the possible ``true'' GEDI footprint center locations, and suggests the ``true'' GEDI footprint location is within 22.5 m in the easting and northing directions from the reported footprint center. 

\begin{figure}[!ht]
\begin{center}
\includegraphics[width=\textwidth]{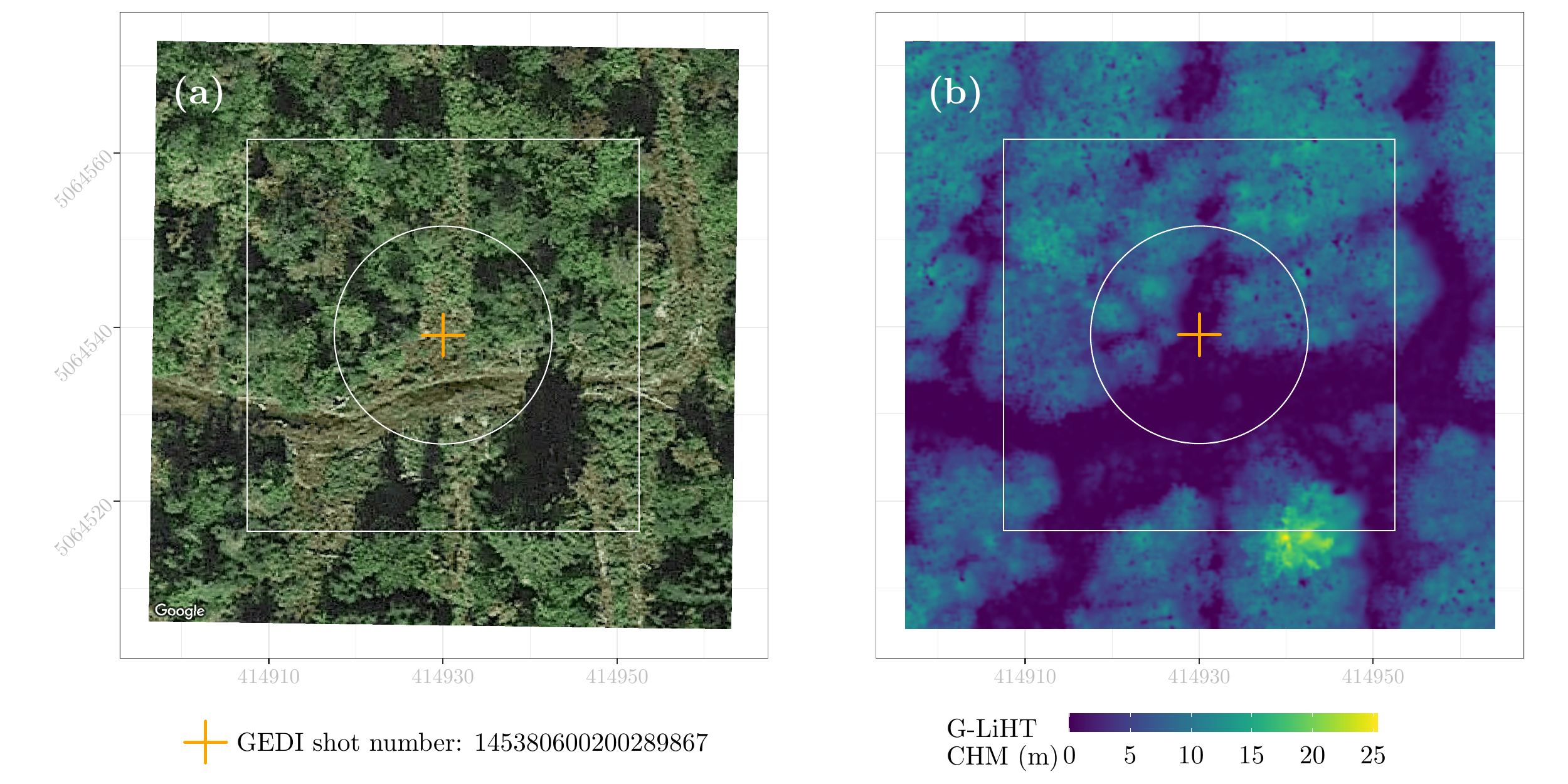}
\end{center}
\caption{(a) Satellite imagery for a single reported GEDI footprint, with 45 m square focal area. The ``GEDI shot number'' is the unique footprint identifier provided with LP DAAC data. (b) G-LiHT canopy height model (CHM) generated from the 3D point cloud measurements for the same focal area displayed in (a). In both (a) and (b) the GEDI footprint extent is shown as a white circle, focal area is a white square, and buffered focal area is the image extent.}
\label{fig:G-LiHT}
\end{figure}

In Figure~\ref{fig:G-LiHT}a, the footprint center for a single GEDI shot is displayed as an orange cross-hairs within the 25 m diameter circular footprint. The square surrounding the GEDI footprint represents the 45 m square focal area (search domain). The total spatial extent of Figure~\ref{fig:G-LiHT}a represents the 70 m square buffer, from which G-LiHT ALS point cloud data can be used to simulate GEDI within the focal area.

\subsubsection{Quality filtering}\label{sec:gedi_filtering}

The Version 2 GEDI L2A data were filtered according to the recommendations outlined in the GEDI Level 2 User Guide \citep{dubayah_et_al_2021_gedi_dataset}. Specifically, GEDI data were filtered to include only measurements taken by the power beams, nighttime acquisitions, and quality flag value of 1. To ensure GEDI data were collected during leaf-on conditions, only GEDI data collected during the 2021 growing season (months of June, July, and August) were considered. Finally, only GEDI footprints with spatially coinciding ALS transects were retained (Figure~\ref{fig:study_area}b). In total, $n$=222 GEDI shots were considered for this study, all of which were collected on July 7, 2021 due to the strict filtering criteria considered here. By only considering high-quality GEDI footprints collected during a single day, effects of temporal variability in the GEDI measurement quality and geolocation error due to factors such as orbital path, cloud cover, atmospheric conditions, and temporal misalignment with the G-LiHT ALS data should be minimized \citep{roy_et_al_2021}. 

\subsection{G-LiHT ALS data}
G-LiHT data were collected over the Maine CMS study area between July 27 and August 6, 2021, along 16 established north-south transects (Figure~\ref{fig:study_area}a)\citep{cohen_et_al_2020}. The G-LiHT data were delivered as individual LiDAR returns in the form of a 3D point cloud, including classified ground returns and vertical measurements interpreted as height above the interpolated ground surface, with a sampling density up to 12 laser pulses/m$^2$. Repeat acquisitions of buildings, road surfaces, and ground beneath forest canopies demonstrated 10 cm (1$\sigma$) accuracy and precision following GPS-INS post-processing and boresight alignment.        Figure~\ref{fig:G-LiHT}b illustrates the G-LiHT point cloud derived canopy height model (CHM) for the example focal area introduced in Section~\ref{sec:gedi}. 

\subsection{Simulation}\label{sec:simulation}
Model development in Section~\ref{sec:models} uses a GEDI RH metric simulator similar to that outlined in \cite{hancock_et_al_2019}. The simulator computes GEDI RH metrics using a Gaussian-weighting of all G-LiHT point cloud measurements within a 25 m radius of the simulated GEDI footprint, with weighting parameters that reflect the GEDI footprint kernel \citep[see Equation 1 in][]{hancock_et_al_2019}. Specifically, for a single simulated footprint centered at spatial coordinates easting $\ell_1$ and northing $\ell_2$, the weight given to the $j$-th 3D G-LiHT point $(x_j, y_j, z_j)$ that falls within the footprint radius is calculated as
\begin{equation}\label{gedi_kernel}
     w_j = \frac{1}{\sigma_f \sqrt{2\pi}} e^{-0.5\frac{(x_j - \ell_1)^2 + (y_j - \ell_2)^2}{\sigma^2_f}},
\end{equation}
where $\sigma_f$ is the Gaussian kernel decay parameter (here, $\sigma_f = 5.5$ m to give $> 97 \%$ of the total weight to the G-LiHT points within 12.5 m of the simulated footprint center) and $x_j$ and $y_j$ specify the easting and northing coordinates, respectively, of the $j$-th G-LiHT point. In this way, G-LiHT points at the center of the simulated footprint are given the greatest weight, with decreasing weight as distance increases from the footprint center. Following methods in \cite{hancock_et_al_2019}, the observed G-LiHT canopy height $z_j$ (m) and associated weight $w_j$ for the $j = \left(1,2,\ldots,h\right)$ G-LiHT points within the simulated GEDI footprint were used to compute RH metrics that approximate well the expected GEDI RH metrics. This simulator is denoted as function $g(\cdot)$ in Section~\ref{sec:models} and is used in the proposed model-based ``search'' for the most likely location(s) from which the observed GEDI RH metrics derive within proximity of the recorded GEDI location. 

\section{Methods}\label{sec:models}
Here we consider models to quantify geolocation error using locations where GEDI and G-LiHT data spatially coincide. Let $\bs_i$ be a spatial coordinate vector with easting $s_{i,1}$ and northing $s_{i,2}$ elements for the $i$-th recorded GEDI footprint center (e.g., coordinates of the orange cross-hairs in Figure~\ref{fig:G-LiHT}b) and $A_i$ be the buffered focal area centered on $\bs_i$ (e.g., the 70-by-70 m square extent of G-LiHT data in Figure~\ref{fig:G-LiHT}b). Prior to model fitting the geographic coordinate system used to index the $n$=222 GEDI observations and associated data (see, e.g., Figure~\ref{fig:study_area}) was converted to a local coordinates system that placed each $\bs_i$ at the center of the $i$-th 70-by-70 m buffered focal area (i.e., each $\bs_i$ was transformed from its original geographic coordinates to $(35, 35)$). Then the posited model for the $i$-th location's GEDI canopy height metric vector is
\begin{align}
     \bz(\bs_i) &= \balpha + \bbeta\odot g\left(\bell(\bs_i),\, \btheta;\; \text{\emph{Data}}(A_i)\right) + \bepsilon(\bs_i),\label{z_model}\\
     \bell(\bs_i) &= \bmu_{\bell} + \bepsilon_{\bell}(\bs_i)\label{l_model},
\end{align}
where $\bz(\bs_i) = (z_1(\bs_i), z_2(\bs_i), \ldots, z_m(\bs_i))^\top$ comprises $m$ canopy metrics (e.g., recorded GEDI RH metrics at $m$ percentiles), $\balpha = (\alpha_1, \alpha_2, \ldots, \alpha_m)^\top$ and $\bbeta = (\beta_1, \beta_2, \ldots, \beta_m)^\top$ are parameter vectors to be estimated, $g(\cdot)$ is a function that returns a vector of length $m$ that approximates $\bz(\bell(\bs_i))$ at geographic location $\bell(\bs_i)=\left(\ell_1(\bs_i), \ell_2(\bs_i)\right)^\top$ using a set of parameters $\btheta$ and high spatial accuracy ALS data expressed as $\text{\emph{Data}}(A_i)$ with area $A_i$ that covers locations $\bs_i$ and $\bell(\bs_i)$ (view $\bell(\bs_i)$ as a coordinate vector at or around $\bs_i$). For our setting, $\bell(\bs_i)$ is to be estimated, $\text{\emph{Data}}(A_i)$ is the G-LiHT data within the buffered focal area used to simulate GEDI RH metrics described in Section~\ref{sec:simulation}, and $\btheta$ holds parameters used within $g(\cdot)$ to map information in $\text{\emph{Data}}(A_i)$ to elements of $\bz(\bell(\bs_i))$. Following the description of $g(\cdot)$ in Section~\ref{sec:simulation}, $\btheta$ holds the GEDI footprint radius and kernel decay parameter $\sigma_f$. The $\odot$ operator denotes component-wise multiplication between the left and right operand vectors. Parameter vectors $\balpha$ and $\bbeta$ provide additive and multiplicative adjustments, respectively, to the $m$ outputs of $g(\cdot)$. The elements of the residual vector $\bepsilon(\bs_i) = (\epsilon_1(\bs_i), \epsilon_2(\bs_i),\ldots,\epsilon_m(\bs_i))^\top$ follow a zero centered normal ($N$) distribution $N\left(0, \tau^2_j\right)$ for $j = (1, 2, \ldots, m)$.


Considering (\ref{l_model}), the coordinate vector $\bell(\bs_i)$ has mean vector $\bmu_{\bell}=(\mu_{\ell,1}, \mu_{\ell,2})^\top$ and error vector $\bepsilon_{\bell}(\bs_i)$ with elements that follow a zero centered normal distribution $N\left(0, \sigma^2_{\ell, k}\right)$ for $k = (1, 2)$. Given the local coordinate system used, $\bmu_{\bell}$ can be interpreted as the average location that minimizes the residual variance between observed canopy height metrics recorded at $\bs_1, \bs_2,\ldots, \bs_n$ and estimated at $\bell(\bs_1), \bell(\bs_2),\ldots, \bell(\bs_n)$.

Model parameters in (\ref{z_model}) and (\ref{l_model}) were estimated using a Bayesian approach, see, e.g., \cite{gelman2013}. Canopy height metrics from the $n$ locations were stacked into a $nm$ length vector $\bz = \left(\bz(\bs_1)^\top, \bz(\bs_2)^\top, \ldots, \bz(\bs_n)^\top\right)^\top$ and parameter input to $g(\cdot)$ was stacked into a $2n$ length vector $\bell = \left(\bell(\bs_1)^\top, \bell(\bs_2)^\top, \ldots, \bell(\bs_n)^\top\right)^\top$. For this analysis we assumed elements in $\btheta$ were known and fixed; however, they could be estimated if desired (see Section~\ref{sec:simulation} for the definition of $g(\cdot)$ and $\btheta$ values). Following (\ref{z_model}) and (\ref{l_model}), the parameters' joint posterior distribution $p\left(\balpha, \bbeta, \btau, \bell, \bmu_{\bell}, \bsigma_{\bell} \given \bz, \btheta, \text{\emph{Data}}\right)$, where $\btau = (\tau^2_1, \tau^2_2, \ldots, \tau^2_m)^\top$,  $\bsigma_{\bell} = (\sigma^2_{\ell,1}, \sigma^2_{\ell,2})^\top$, and $\text{\emph{Data}}$ holds all G-LiHT data within $A_1, A_2,\ldots,A_n$, is proportional to the likelihood times the prior distributions 
\begin{align}\label{full_ltd}
     \prod^n_{i=1}&\prod^m_{j=1}\underbracket[1pt]{N\left(z_j(\bs_i) \given \alpha_j + \beta_j g_j\left(\bell(\bs_i); \btheta, \text{\emph{Data}}(A_i)\right), \tau^2_j\right)}_\text{Likelihood}\times\nonumber\\
     &\prod^m_{j = 1}N\left(\alpha_j \given \mu_{\alpha}, \sigma^2_{\alpha}\right)\times \prod^m_{j = 1}N\left(\beta_j \given \mu_{\beta}, \sigma^2_{\beta}\right) \times \prod^m_{j=1}IG\left(\tau^2_j \given a_{\tau}, b_{\tau}\right) \times\\
     &\prod^n_{i=1}\prod^2_{k=1}N\left(\ell_k(\bs_i) \given \mu_{\ell,k}, \sigma^2_{\ell,k}\right)\times \prod^2_{k=1}N\left(\mu_{\ell,k} \given s_{k}, \sigma^2_{\mu_{\ell},k}\right) \times \prod^2_{k=1} IG\left(\sigma^2_{\ell,k} \given a_\ell, b_\ell\right).\nonumber
\end{align}
Hyperparameters for the prior distributions in (\ref{full_ltd}) were chosen to be weakly informative. Specifically, we assigned a normal prior distribution to elements of the likelihood's additive adjustment vector $\balpha$ with mean $\mu_\alpha=0$ and variance $\sigma^2_\alpha = \text{1,000}$. Similarly, elements in the multiplicative adjustment vector $\bbeta$ are assigned a normal prior distribution with mean $\mu_\beta = 1$ and variance $\sigma_\beta^2=$ 1,000. Notice, prior means of 0 and 1 for elements of $\balpha$ and $\bbeta$, respectively, places some weight on the prior belief that the output of $g(\cdot)$ does not need additive or multiplicative adjustment, although the large variances allow for substantial learning from the data. Elements in the likelihood's variance parameter vector $\btau$ each follow an inverse-Gamma ($IG$) prior distribution with shape and scale hyperparameters $a_\tau = 2$ and $b_\tau = 10$, respectively. For the $IG$, when the shape equals 2, the distribution's mean equals the scale and its variance is infinite. Here, we chose the scale value from exploratory analysis using linear regression defined by the simplified likelihood $\prod^n_{i = 1}\prod^m_{j=1}N(z_j(\bs_i) \given g_j(\bs_i; \btheta, \text{\emph{Data}}(A_i)), \tau^2_j)$ (i.e., assuming no geolocation error nor additive and multiplicative adjustment). The induced prior on the easting and northing components in $\bell(\bs_i)$ are normal with mean $\mu_{\ell,1}$ and $\mu_{\ell,2}$ and variances $\sigma^2_{\ell,1}$ and $\sigma^2_{\ell,2}$, respectively. We set the hyperprior distributions for $\mu_{\ell,1}$ and $\mu_{\ell,2}$ (i.e., elements of $\bmu_{\bell}$) as normal with means $s_{1}$=35 and $s_{2}$=35 and variances $\sigma^2_{\mu_{\ell},1} = \sigma^2_{\mu_{\ell},2} =  \text{1,000}$. Notice, given the local coordinate system, $\bs_i = (s_{i,1}, s_{i,2})$ equal (35, 35) for all $n$ focal areas. Said differently, we are setting the prior mean for $\bell(\bs_i)$ equal to $\bs_i$ to reflect prior belief that the reported location $\bs_i$ for the GEDI footprint is accurate, but allowing a large variance so the data can inform the estimate of $\bell(\bs_i)$. Lastly, hyperprior distributions for $\sigma^2_{\ell,1}$ and $\sigma^2_{\ell,2}$ are $IG$ with $a_\ell = 2$ and $b_\ell = 100$. This choice of $IG$ scale expresses prior belief the distribution of all $\bell(\bs_i)$ has a standard deviation of 10 m about $\bmu_{\bell}$; however, between the large values set for $\sigma^2_{\mu_{\ell},1}$ and $\sigma^2_{\mu_{\ell},2}$ as well as the uninformative specification induced by $a_\ell$, the data far outweigh the prior.

In application, we bound support for elements of $\bell(\bs_i)$ to be within 22.5 m of $\bs_i$. This bounding restricts the area within which the model can search for the ``true'' location from which $\bz(\bs_i)$ was measured (i.e., the search for $\bell(\bs_i)$ occurs within the white focal area box depicted in Figure~\ref{fig:G-LiHT}). The 22.5 m bounding was used because it represents the sum of the estimated geolocation error of GEDI (10 m) and the GEDI footprint radius (12.5 m). 

\subsection{Submodel}\label{sec:submodel}

While (\ref{z_model}) subsumes many submodels, some of which are discussed in Section~\ref{sec:discussion}, the one we consider here is a model-based analog to several GEDI geolocation correction approaches, see, e.g., \cite{quiros_et_al_2021}, \cite{roy_et_al_2021}, and, \cite{blair_and_hoften_1999}. Again using the local coordinate system across all $n$ observed locations, we estimate $\bell^{\ast}=\left(\ell^{\ast}_1, \ell^{\ast}_2\right)^\top$ defined as the single coordinate vector that minimizes residual variance between the observed $\bz(\bs_i)$ and the additive and multiplicative adjusted $g\left(\bell^{\ast}; \btheta, \emph{Data}(A_i)\right)$ for $i=(1,2,\ldots,n)$. The posited submodel for the $i$-th location is then
\begin{equation}\label{z_submodel}
     \bz(\bs_i) = \balpha + \bbeta\odot g\left(\bell^{\ast}; \btheta, Data(A_i)\right) + \bepsilon(\bs_i),
\end{equation}
where parameters $\balpha$, $\bbeta$, and $\bepsilon(\bs_i)$ were defined previously for (\ref{z_model}). The parameters' joint posterior distribution $p\left(\balpha, \bbeta, \btau \given \bz, \btheta, \text{\emph{Data}}\right)$ is proportional to
\begin{align}\label{sub_ltd}
     \prod^n_{i=1}&\prod^m_{j=1}\underbracket[1pt]{N\left(z_j(\bs_i) \given \alpha_j + \beta_j g_j\left(\bell^\ast; \btheta, \text{\emph{Data}}(A_i)\right), \tau^2_j\right)}_\text{Likelihood}\times\nonumber\\
     &\prod^m_{j = 1}N\left(\alpha_j \given \mu_{\alpha}, \sigma^2_{\alpha}\right)\times \prod^m_{j = 1}N\left(\beta_j \given \mu_{\beta}, \sigma^2_{\beta}\right) \times \prod^m_{j=1}IG\left(\tau^2_j \given a_{\tau}, b_{\tau}\right) \times
     \prod^2_{k=1}N\left(\ell^{\ast}_k \given s_k, \sigma^2_{\ell^{\ast},k}\right).
\end{align}
All prior distributions, with the exception of those for elements of $\bell^{\ast}$, were defined previously. The prior distribution for $\ell^{\ast}_1$ and $\ell^{\ast}_2$ are normal with means $s_1$=35 and $s_2$=35, respectively, and variances $\sigma^2_{\ell^{\ast},1} = \sigma^2_{\ell^{\ast},2} = $ 1,000. Again, in application, we impose bounded support for $\ell^{\ast}_1$ and $\ell^{\ast}_2$ to be within 22.5 m of $(35,35)$. 


\subsection{Implementation, posterior summaries, and fitted values}\label{sec:implementation}
Models were implemented in C++ and called Fortran \texttt{openBLAS} \citep{zhang13} and Linear Algebra Package (LAPACK; www.netlib.org/lapack) libraries for efficient matrix computations. \texttt{openBLAS} is an implementation of Basic Linear Algebra Subprograms (BLAS; www.netlib.org/blas) capable of exploiting multiple processors. (\emph{All code and data will be provided on a repository to allow for reproduction of results. The code and data may be hosted by the publishing journal or public archive, and provided prior to that if requested by reviewers.})

We refer to (\ref{z_model}) and (\ref{z_submodel}) as the \emph{full model} and \emph{submodel}, respectively. Inference for these models is based on samples from parameters' posterior distributions. These samples were collected via a Markov chain Monte Carlo (MCMC) algorithm that used a Gibbs sampler for those parameters with closed form full conditional distributions (i.e., $\bbeta$, $\balpha$, $\btau$, $\bmu_{\bell}$ and $\bsigma_{\bell}$) and Metropolis samplers for all other parameters \citep[see, e.g.,][for a general description of these samplers]{gelman2013}. While developing the sampler for $\bell$, we found the posterior surface for a given $\bell(\bs_i)$ could be highly multimodal. The posterior mode for $\bell(\bs_i)$ is a location at or around $\bs_i$ that minimizes the log of the density given in (\ref{full_ltd}), and hence represents a location that likely generated the observed $\bz(\bs_i)$. Multiple modes were observed when more than one location at or around $\bs_i$ yielded comparable values for (\ref{full_ltd}) or when local modes were present. Such posterior distributions present a challenge for the basic Metropolis sampler because the MCMC chain tends to get ``stuck'' in local modes and fails to adequately explore the parameter space. Our solution was to use the repelling–attracting Metropolis (RAM) algorithm developed by \cite{tak_et_al_2018} that maintains the computational advantages and simple implementation of the Metropolis algorithm but allows for improved sampling from multimodal posterior distributions.    

The MCMC algorithm provides samples from parameters' posterior distributions. Posterior inference reported in Section~\ref{sec:results} is based on $M$ = 50,000 post-convergence and thinned samples from five MCMC chains, i.e., 10,000 from each chain. We used convergence diagnostics and thinning rules outlined in \cite{gelman2013}. Point and interval estimates for $\balpha$, $\bbeta$, and variance parameters presented in Section~\ref{sec:results} include posterior medians and 95\% credible intervals.

Our particular interest is in posterior inferences about the elements of $\bell$ because they may reveal the likely ``true'' location from which a given $\bz(\bs_i)$ was measured and systematic geolocation errors across the focal areas. In addition to $\bmu_{\bell}$, which is directly estimated by the full model, we estimate a maximum a posteriori probability (MAP) for each $\bell(\bs_i)$, denoted as $\bseta_{\bell}(\bs_i)$, and a composite MAP estimate over all $n$ focal areas, denoted as $\bseta_{\bell}$. Here, $\bseta_{\bell}(\bs_i)$ is the posterior distribution's largest valued mode within the $i$-th focal area, and $\bseta_{\bell}$ is the largest valued mode accumulated over all $n$ $\bell(\bs_i)$ posterior distributions (i.e., the mode of the composite posterior distribution formed using all posterior distribution samples from $\bell(\bs_i)$ for $i=(1,2,\ldots,n)$). Similarly, for the submodel we estimate $\bseta_{\bell^\ast}$ which is the largest valued mode of $\bell^\ast$'s posterior distribution. 

\begin{figure}[!ht]
\begin{center}
\includegraphics[width=\textwidth]{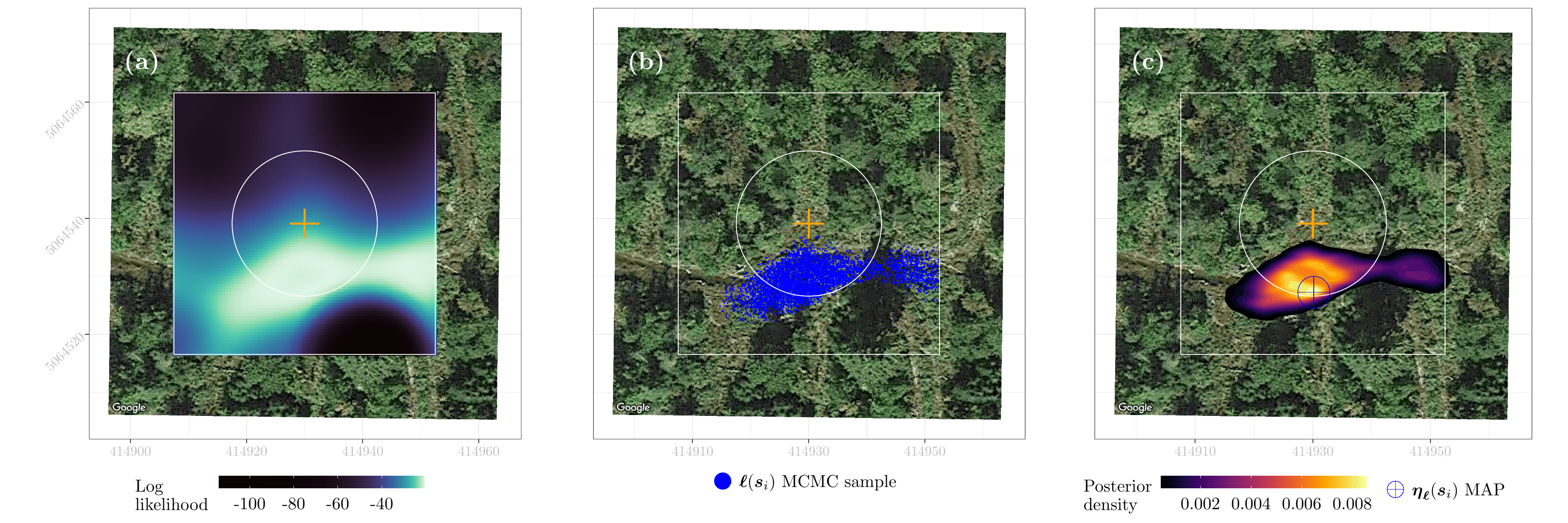}
\end{center}
\caption{Continued illustration of focal area $i$=176 used in Figure~\ref{fig:G-LiHT}. (a) Log likelihood surface calculated using parameter point estimates from the full model. (b) Full model's MCMC samples from $\bell(\bs_i)$ posterior distribution, where $\bs_i$ is the given GEDI footprint's center. (c) Surface plot reflecting the density of MCMC samples shown in (b), used to better visualize $\bell(\bs_i)$'s posterior distribution and key features such as the posterior's maximum a posteriori probability (MAP) estimate.}\label{fig:example_plots}
\end{figure}

MCMC samples from the full model's $\bell(\bs_i)$ define coordinates within the $i$-th focal area and hence can be visualized as locations on a map. The density of these samples is greater at and around locations that minimize the log of the density given in (\ref{full_ltd}). This visualization is illustrated in Figure~\ref{fig:example_plots} for GEDI focal area $i$=176 (i.e., similar figures for all $n$ locations are given in the Supplementary Material Figures~S2-S223). Figure~\ref{fig:example_plots}a shows the log likelihood surface that is generated by plugging in posterior mean point estimates for $\balpha$, $\bbeta$, and $\btau$ and evaluating the log likelihood at 10,000 grid locations within the focal area (i.e., each evaluation consists of setting $\bell(\bs_{176})$ to a grid cell location and then computing the log likelihood). Importantly, we compute this likelihood surface only for illustrative purposes and as an ``agreement check'' for the MCMC-based $\bell(\bs_{176})$ estimates (it is used for no other purpose in the analysis). Looking at this likelihood surface, we expect more posterior  $\bell(\bs_{176})$'s samples (i.e., sampled at a higher frequency) in regions of high log likelihood. Figure~\ref{fig:example_plots}b shows a subset of 5,000 MCMC samples from $\bell(\bs_{176})$'s posterior distribution and that sampling density generally follows the log likelihood surface in Figure~\ref{fig:example_plots}a. Using these $\bell(\bs_{176})$ posterior samples, we can compute point and interval estimates of the easting $\ell_1(\bs_{176})$ and northing $\ell_2(\bs_{176})$ posteriors. While such summaries of individual axes are potentially useful, we are mainly interested in the distance and direction between the recorded GEDI location $\bs_{176}$ and $\bell(\bs_{176})$'s mode(s). Given MCMC samples from $\bell(\bs_{176})$'s posterior, e.g., Figure~\ref{fig:example_plots}b, we use a two-dimensional kernel density estimation algorithm described by \cite{venables_ripley_2002} and implemented in the \texttt{R} \texttt{MASS} package to visualize $\bell(\bs_{176})$'s posterior distribution and estimate its MAP location. Figure~\ref{fig:example_plots}c shows the resulting posterior density image and estimated MAP $\bseta(\bs_{176})$. In this way, the posterior density images and associated $\bseta_{\bell}(\bs_i)$ for each of the $n$ focal areas and composite MAP estimate for $\bseta_{\bell}$ and submodel's $\bseta_{\bell^\ast}$ were computed.

Given MCMC samples from the full model's $\bell(\bs_i)$, we generate samples from two additional posterior distributions of interest: 1) $d_{\bell}(\bs_i)$ distance between the $i$-th focal area's $\bell(\bs_i)$ and $\bs_i$, and 2) $d_{\bell}$ distance between all $\bell(\bs_i)$ for $i=(1,2,\ldots,n)$ and $(35, 35)$. Posterior samples for $d_{\bell}(\bs_i)$ are generated using composition sampling (i.e., one sample from $\bell(\bs_i)$ yields one sample from $d_{\bell}(\bs_i)$) via
\begin{align*}
d_{\bell}^{(l)}(\bs_i) &= \left[\left(\ell_1^{(l)}(\bs_i) - s_{i,1}\right)^2+\left(\ell_2^{(l)}(\bs_i) - s_{i,2}\right)^2 \right]^{0.5}\\
&= \lVert \bell^{(l)}(\bs_i) - \bs_i \rVert \text{ for } l = (1, 2, \ldots, M),
\end{align*}
where the superscript $l$ in parentheses indexes MCMC samples. Similarly, samples for $d_{\bell}$ are the collection of all focal area samples $\left(d_{\bell}^{(l)}(\bs_1), d_{\bell}^{(l)}(\bs_2), \ldots, d_{\bell}^{(l)}(\bs_n)\right)$ for $l = (1, 2, \ldots, M)$. Samples from these parameters' posterior distributions allow us to explore within and across focal area differences between $\bz(\bs_i)$'s reported and estimated location.

If a given $d_{\bell}(\bs_i)$'s posterior distribution has a central tendency at or very near to zero, then there is evidence in support of no geolocation error. If $d_{\bell}$'s posterior distribution has central tendency at or very near to zero, then there is evidence in support of no systematic error across locations. For either parameter, a posterior central tendency away from zero suggests geolocation error. A broad flat posterior distribution that covers zero and values far from zero suggests inconclusive results. As we will see in Section~\ref{sec:results}, such cases often occur when the focal area's vertical vegetation structure is homogeneous. In such settings, many locations might, or might not, match the observed outcome $\bz(\bs_i)$ well, with the result being a broad posterior distribution that does not identify distinctly likely locations from which $\bz(\bs_i)$ was measured. 

MCMC samples from the submodel's $\bell^\ast$ are used to generate samples from the posterior distribution of the distance between $\bell^\ast$ and $(35,35)$, denoted as $d_{\bell^{\ast}}$. Again, the desired samples are collected using composition sampling via 
\begin{equation}
d_{\bell^{\ast}}^{(l)} = \lVert \bell^{\ast,(l)}- (35,35)^\top \rVert \text{ for } l = (1, 2, \ldots, M).
\end{equation}

Like $d_{\bell}$, the $d_{\bell^{\ast}}$ parameter is a measure of systematic geolocation error (i.e., bias) across the $n$ focal areas. If $d_{\bell^{\ast}}$'s posterior distribution has central tendency away from zero, then we can conclude there is systematic geolocation error. If $d_{\bell^{\ast}}$'s posterior distribution has central tendency at or very near to zero, then we can conclude there is no systematic geolocation error, but cannot say anything about geolocation error at any given focal area. We must look to the full model's $\bell(\bs_i)$ to quantify non-systematic geolocation error at each focal area.

Posterior distributions for the angle between $\bs_i$ and $\bell(\bs_i)$ and $\bell^\ast$ are computed via composition sampling analogous to each $d_{\bell}(\bs_i)$ and $d_{\bell^\ast}$, and are useful for describing patterns in directional geolocation error. 

We are also interested in quantifying the distances between $\bs_i$ and the full model's focal area specific MAP $\bseta_{\bell}(\bs_i)$, i.e., $d_{\bseta_{\bell}}(\bs_i) = \lVert\bseta_{\bell}(\bs_i) - \bs_i\rVert$ for $i = (1, 2, \ldots, n)$, and cumulative MAP $\bseta_{\bell}$, i.e., $d_{\bseta_{\bell}} = \lVert\bseta_{\bell} - \bs\rVert$. We may also compute the direction between $\bs_i$ and each $\bseta_{\bell}(\bs_i)$ and $\bseta_{\bell}$.

Estimating the posterior distribution for model fitted values is a key step in assessing model fit to the observed data and geolocation error extent. Given MCMC samples from model parameters' posterior distributions, we again use composition sampling to collect samples from the model fitted value's posterior distribution. For the full model, $M$ samples from the fitted value's posterior distribution for the $j$-th outcome at the $i$-th focal area are collected via 
\begin{equation}\label{z_fitted}
     \hat{z}^{(l)}_j(\bs_i) = N\left(\alpha^{(l)}_j + \beta^{(l)}_jg_j\left(\bell^{(l)}(\bs_i); \btheta,  Data(A_i)\right), \tau^{2,(l)}_j\right) \text{ for } l = (1,2,\ldots, M).
\end{equation}
We then compute the posterior distribution median over the $M$ MCMC samples for each of the $m$ outcomes and $n$ locations. The vector of median estimates is $\hat{\bz} = (\hat{\bz}(\bs_1)^\top, \hat{\bz}(\bs_2)^\top, \ldots, \hat{\bz}(\bs_n)^\top)^\top$, where $\hat{\bz}(\bs_i) = (\hat{z}_1(\bs_i), \hat{z}_2(\bs_i), \ldots, \hat{z}_m(\bs_i))^\top$. 
Similarly, submodel fitted values are collected using its parameters' posterior samples via
\begin{equation}\label{z_submodel_fitted}
     \hat{z}^{\ast,(l)}_j(\bs_i) = N\left(\alpha^{(l)}_j + \beta^{(l)}_jg_j\left(\bell^{\ast,(l)}; \btheta, Data(A_i)\right), \tau^{2,(l)}_j\right) \text{ for } l = (1,2,\ldots, M),
\end{equation}
where for all $m$ outcomes and $n$ focal areas, the resulting posterior distribution median estimates $\hat{\bz}^\ast$ are stacked analogous to $\hat{\bz}$.

\begin{equation}\label{z_center_fitted}
     \dot{z}_j(\bs_i) = g_j\left(\bs_i ;\btheta, Data(A_i)\right).
\end{equation}
Here, if the simulator is a good approximation for GEDI and there is no geolocation error, then $\dot{\bz} = (\dot{\bz}(\bs_1)^\top, \dot{\bz}(\bs_2)^\top, \ldots, \dot{\bz}(\bs_n)^\top)^\top$, where $\dot{\bz}(\bs_i) = (\dot{z}_1(\bs_i), \dot{z}_2(\bs_i), \ldots, \dot{z}_m(\bs_i))^\top$, should approximate $\bz$ well. If we see $\hat{\bz}$ and/or $\hat{\bz}^\ast$ approximate $\bz$ better than $\dot{\bz}$ then we might conclude there is some level of geolocation error. We use the root mean squared error (RMSE) between the observed $\bz$ and estimated $\dot{\bz}$, $\hat{\bz}^\ast$, and $\hat{\bz}$ for each of the $m$ outcomes to assess location-based fit.  

\section{Results}\label{sec:results}

The two models detailed in Section~\ref{sec:models} (the full model and submodel) were fit to the data described in Section~\ref{sec:data}. We considered $m=11$ RH metrics, specifically RH percentiles starting at 50 and incrementing by 5 to 95, and also 98, which we took for the canopy top height. 

\begin{figure}[!ht]
\begin{center}
\includegraphics[width=14cm,trim={0cm 5.5cm 0cm 5.5cm},clip]{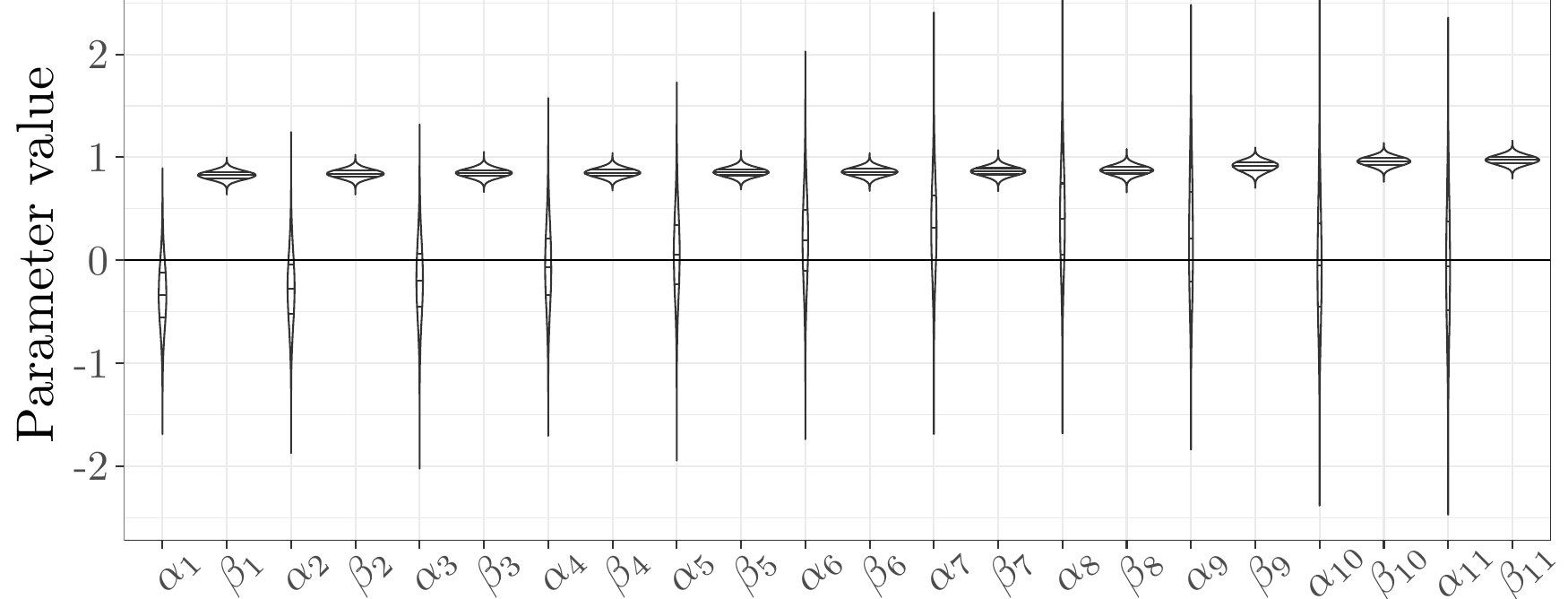}
\end{center}
\caption{Posterior distribution summaries for the full model's $\balpha$ and $\bbeta$. The subscripts $1, 2, \ldots, m$, where $m = 11$, correspond to relative height percentiles starting at 50 and incrementing by 5 to 95, and also 98. The upper and lower horizontal lines within each violin plot represent the posterior distribution's 95\% credible bounds, and the middle horizontal line is the posterior distribution's median. The posterior distribution range is given by the violin plot's extremes. }\label{fig:full_beta_alpha}
\end{figure}

Estimates of $\balpha$ and $\bbeta$ for the full model are given in Figure~\ref{fig:full_beta_alpha}. Following (\ref{z_model}), if there was no geolocation error and the simulator $g(\cdot)$ approximated $\bz$ well, then we would expect elements in $\balpha$ and $\bbeta$ to be 0 and 1, respectively. Considering results given in Figure~\ref{fig:full_beta_alpha}, we can conclude the simulator required only marginal additive and multiplicative adjustments for the lower RH metrics. The two largest RH metrics, i.e., the $95^{\text{th}}$ and $98^{\text{th}}$ percentiles, corresponding estimates of $\alpha_{10}$ and $\alpha_{11}$, respectively, were indistinguishable from 0, and $\beta_{10}$ and $\beta_{11}$, respectively, were indistinguishable from 1. Estimates of $\balpha$ and $\bbeta$ for the submodel were similar to the full model; however, additive adjustments increased with increasing RH metrics (see, Supplementary Material Figure~S1).

\begin{figure}[!ht]
\begin{center}
\includegraphics[width=14cm,trim={0cm 5.5cm 0cm 5.5cm},clip]{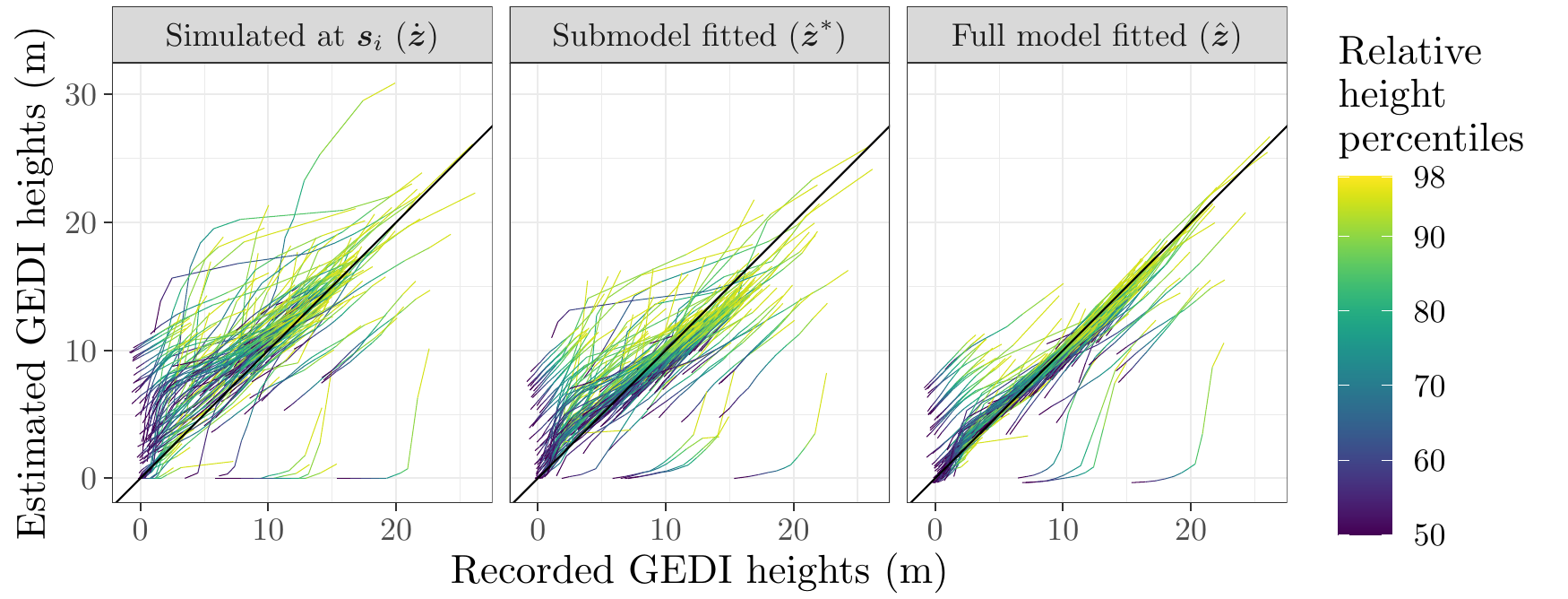}
\end{center}
\caption{Each shaded line connects the $m$=11 relative height percentiles for each $i = (1,2,\ldots, n$) focal areas. A black 1-to-1 line is given in each subplot. Subplots from left to right are: observed versus simulated values at the reported GEDI footprint center; observed versus submodel fitted values, and; observed versus full model fitted values. }\label{fig:obs_vs_fitted}
\end{figure}

\begin{figure}[!ht]
\begin{center}
\includegraphics[width=14cm,trim={0cm 4.2cm 0cm 4.2cm},clip]{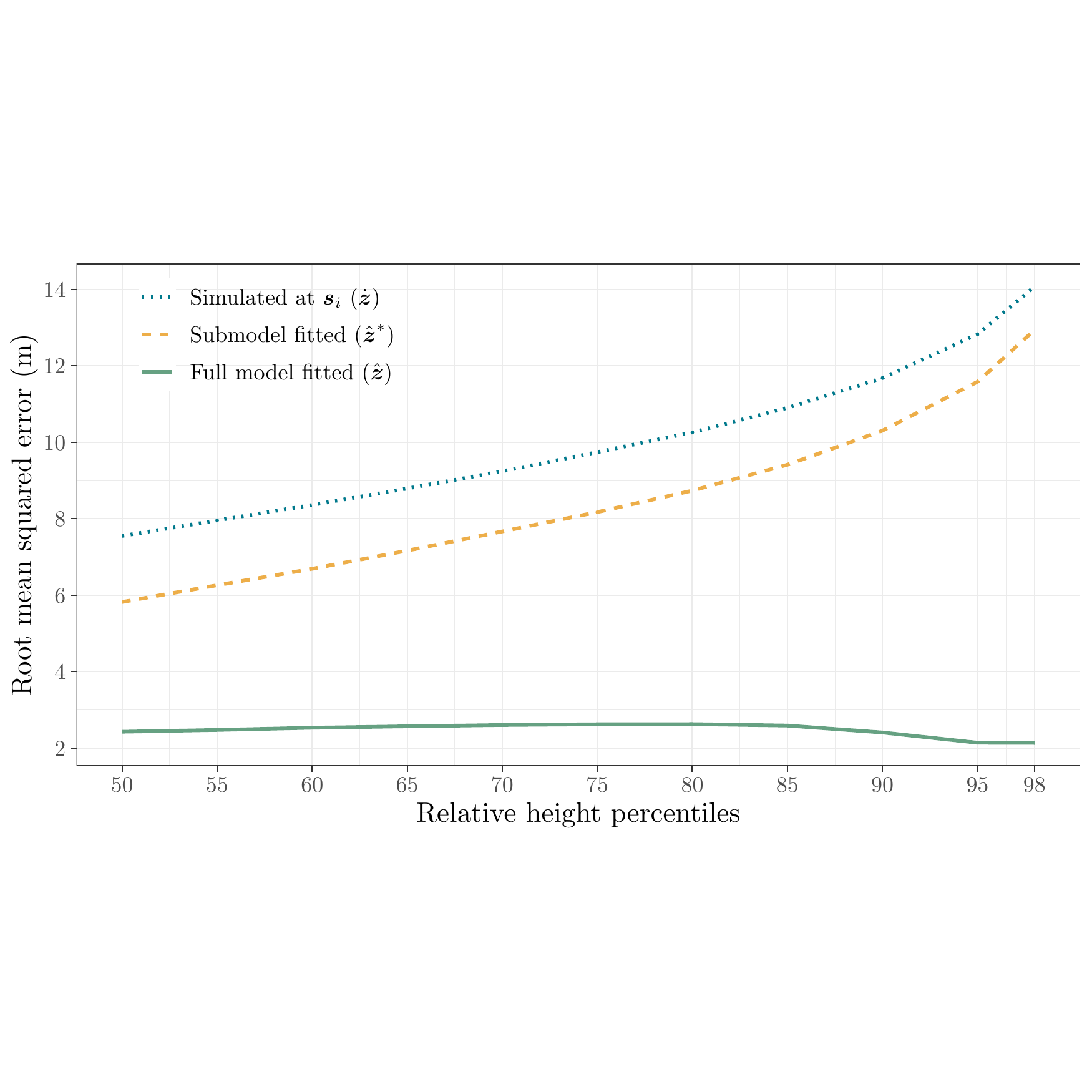}
\end{center}
\caption{Root mean squared error (RMSE) between observed and fitted values. RMSE computed for each RH metric across $n$ = 222 focal areas. These RMSE values summarize the observed versus fitted values shown in Figure~\ref{fig:obs_vs_fitted}.}\label{fig:rmse_obs_vs_fitted}
\end{figure}

Model fitted values are given in Figure~\ref{fig:obs_vs_fitted}. Here, the left subplot shows observed RH values $\bz(\bs_i)$ versus $\dot{\bz}(\bs_i)$ for $i = (1,2,\ldots,n)$ (i.e., simulated using $g(\cdot)$ at the reported GEDI footprint center). As described toward the end of Section~\ref{sec:implementation}, if the simulator provides a good approximation for GEDI and there is no geolocation error, then $\dot{\bz}(\bs_i)$ should provide a good match to $\bz(\bs_i)$. Given the shaded lines in the left subplot do not fall along the 1-to-1 line, we can conclude that either the simulator does not approximate GEDI well or there is geolocation error. It is useful to consider Figure~\ref{fig:rmse_obs_vs_fitted}, which plots RMSE between observed and fitted values by percentile, simultaneously with Figure~\ref{fig:obs_vs_fitted}. As shown by the dotted line in Figure~\ref{fig:rmse_obs_vs_fitted}, the spread of observed versus $\dot{\bz}(\bs_i)$ fitted values in Figure~\ref{fig:obs_vs_fitted} translates to relatively large RMSE values that tend to increase with increasing RH. 

\begin{figure}[!ht]
\begin{center}
    \subfigure[]{\includegraphics[height=6.6cm,trim={0cm 0cm 0cm 0cm},clip]{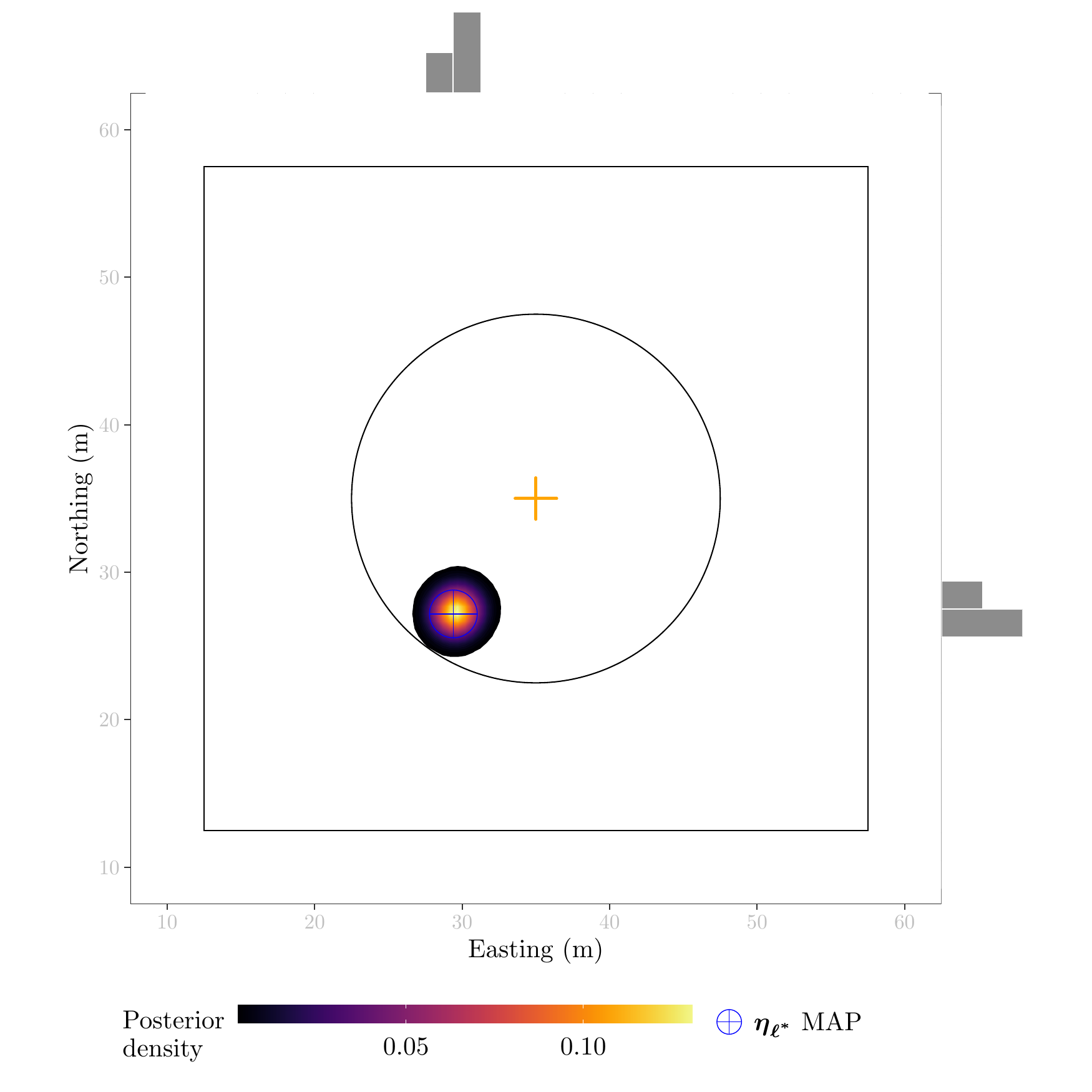}\label{fig:ell_submodel}} 	\subfigure[]{\includegraphics[height=6.6cm,trim={0cm 0cm 0cm 0cm},clip]{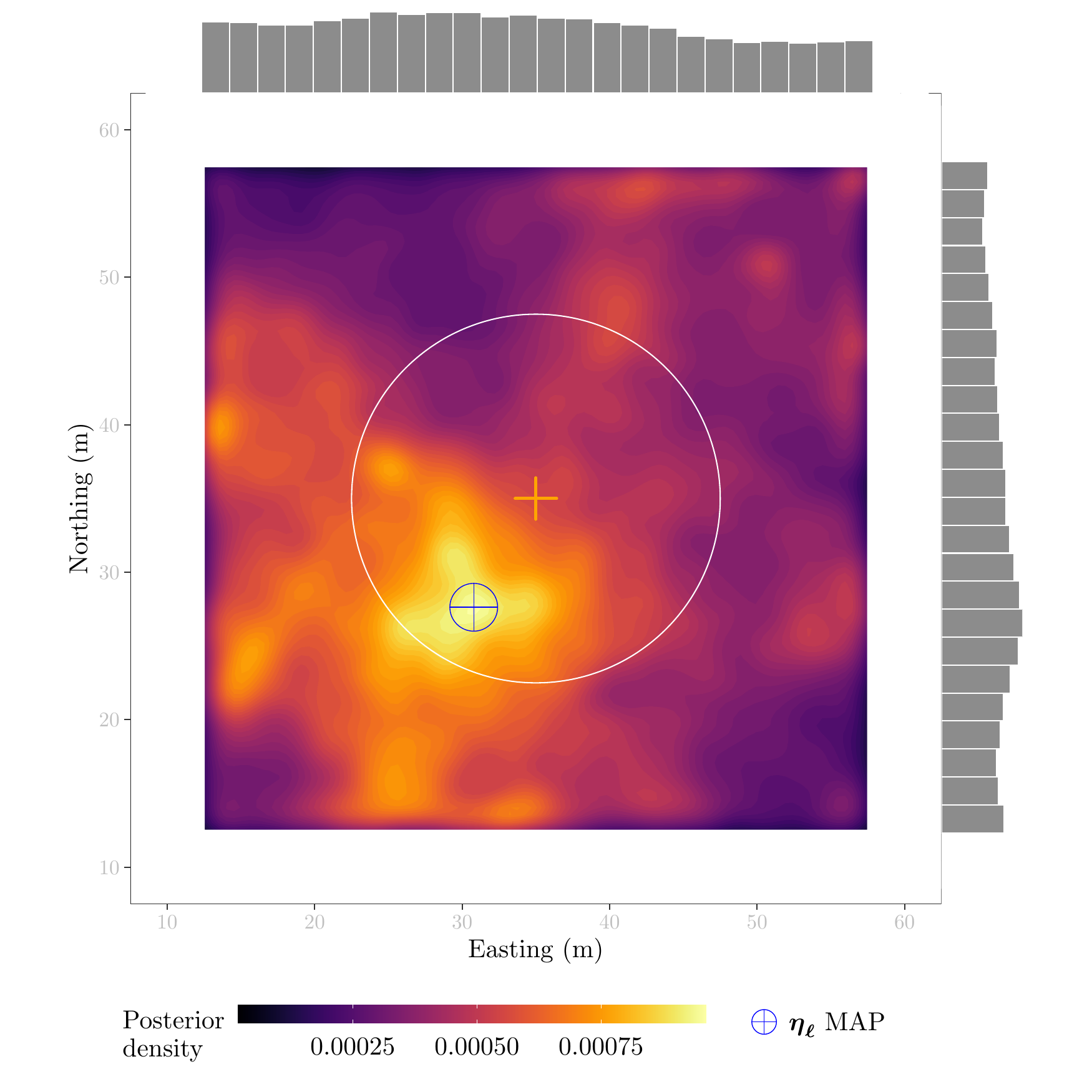}\label{fig:ells_fullmodel}} 
\caption{\subref{fig:ell_submodel} Estimated posterior distribution surface of the submodel's $\bell^{\ast}$. Histograms of $\ell_1^\ast$ and $\ell_2^\ast$ marginal posterior distributions are plotted on the top and right axes, respectively. \subref{fig:ells_fullmodel} Estimated cumulative posterior distribution surface of the full model's $\bell(\bs_i)$ from $i = (1,2, \ldots, n)$. Histograms of $\ell_1$'s and $\ell_2$'s marginal posterior distributions are plotted on the top and right axes, respectively.}\label{fig:post_dist_ells}
\end{center}
\end{figure}

The middle subplot in Figure~\ref{fig:obs_vs_fitted}, and dashed line in Figure~\ref{fig:rmse_obs_vs_fitted}, show that simulating GEDI at the submodel's estimated $\bell^\ast$ improves fit. The estimated posterior distribution for $\bell^\ast$ is illustrated in Figure~\ref{fig:ell_submodel}. The posterior distribution median and 95\% credible interval for elements of $\bell^\ast$ are 29.57 (29.03, 30.40) and 27.27 (26.53, 28.15) for $\ell_1^\ast$ and $\ell_2^\ast$, respectively. Given neither of these credible intervals overlap 35 (i.e., where 35 is the local coordinate system center coordinate and reported GEDI footprint center) we can conclude there is a substantial degree of systematic geolocation error and have a high degree of confidence the systematic error falls within the noted easting and northing credible intervals. The corresponding median and 95\% credible interval estimates for $d_{\bell^\ast}$ are 9.45 (8.43, 10.17) m. The estimated posterior median and 95\% credible interval for the angle between $(35,35)$ and $\bell^\ast$---computed using composition sampling analogous to $d_{\bell^\ast}$---are 234.80$^{\circ}$ (230.98$^{\circ}$, 238.70$^{\circ}$). As shown on Figure~\ref{fig:ell_submodel}, the estimated MAP $\bseta_{\bell^{\ast}}$ is 29.40 easting and 27.17 northing, which is 9.62 m from the observed GEDI footprint center $(35,35)$ at 234.43$^{\circ}$. 

The right subplot in Figure~\ref{fig:obs_vs_fitted}, and solid line in Figure~\ref{fig:rmse_obs_vs_fitted} show that simulating GEDI at the full model's focal area specific estimate of $\bell(\bs_i)$ provides the best fit, relative to the reported location and submodel. The full model's relatively tight fitted value scatter about the 1-to-1 line in Figure~\ref{fig:obs_vs_fitted}, which yields the consistently small RMSE seen in Figure~\ref{fig:rmse_obs_vs_fitted}. 

The posterior distribution median and 95\% credible interval for elements of $\bmu_{\bell}$ are 33.41 (26.77, 41.09) and 33.37 (27.49, 40.14) for $\mu_{\ell,1}$ and $\mu_{\ell,2}$, respectively. These estimates suggest there is no evidence of systematic geolocation error across the $n$ focal areas (i.e., both credible intervals include 35) and, hence, runs counter to the submodel results. However, $\bell$'s posterior surface and its corresponding MAP $\bseta_{\bell}$ location shown in Figure~\ref{fig:ells_fullmodel} support the submodel result of a systematic geolocation error in the southwest direction. Specifically, the estimated MAP $\bseta_{\bell}$ is 30.79 easting and 27.98 northing, which agrees well with the submodel $\bseta_{\bell^\ast}$ estimate. This leads us to look at the interpretation of $\bmu_{\bell}$ more closely. Figure~\ref{fig:ells_fullmodel} shows why $\bmu_{\bell}$ is not a reliable summary of systematic error for these data. Here, histograms of $\ell_1$'s and $\ell_2$'s marginal posterior distributions plotted on the top and right axes of Figure~\ref{fig:ells_fullmodel}, respectively, suggest substantial non-normality. These broad distributions with substantial skew cause estimates of $\bmu_{\bell}$ to be drawn toward the focal area center and have large variability. In contrast, $\bseta_{\bell}$ is simply the largest value on $\bell$'s posterior distribution and hence not affected by deviations from normality.

Estimates of the submodel $\bell^{\ast}$ and $\bseta_{\bell^\ast}$ and full model $\bseta_{\bell}$ are comparable to the geolocation error of the Version 2 GEDI data reported by \cite{dubayah_et_al_2020} (i.e., approximately 10 m). However, the posterior surface in Figure~\ref{fig:ells_fullmodel} hints at geolocation error beyond this limited systematic offset, and suggests substantial geolocation error variability within and across focal areas.  

When quantifying geolocation error, it is useful to distinguish between systematic error which, as defined here, is a summary of bias, and actual error which is the distance between the reported and actual measurement location. We do not have a parameter that summarizes actual geolocation error; however, we can begin to explore beyond systematic geolocation error using $d_{\bell}$ across the $n$ focal areas, and $d_{\bell}(\bs_i)$ and $d_{\bseta_{\bell}}(\bs_i)$ within the $i$-th focal area.

\begin{figure}[!ht]
\begin{center}
    \subfigure[]{\includegraphics[height=6.6cm,trim={0cm 0cm 0cm 0cm},clip]{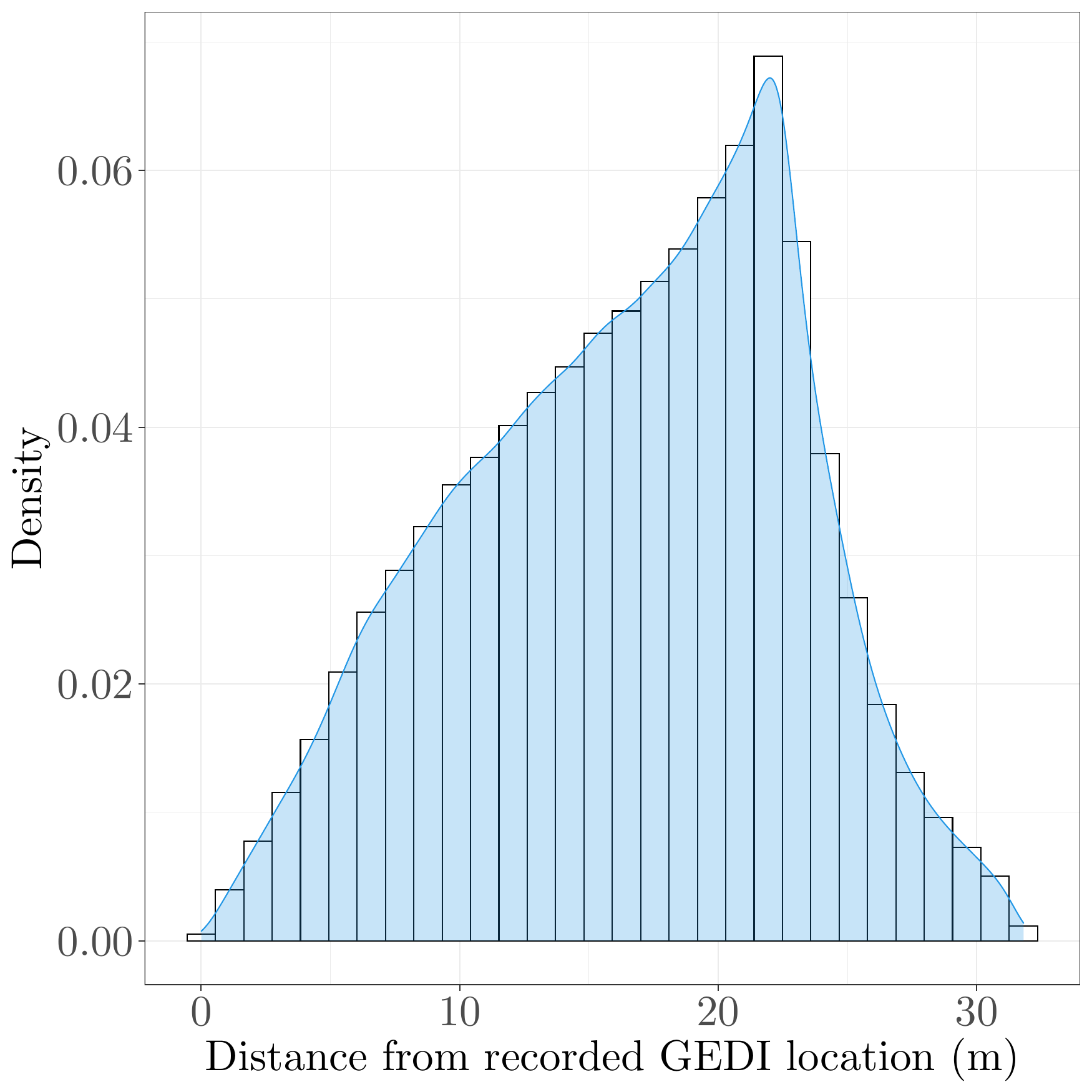} \label{fig:dist_hist}}
	\subfigure[]{\includegraphics[height=6.6cm,trim={0cm 0cm 0cm 0cm},clip]{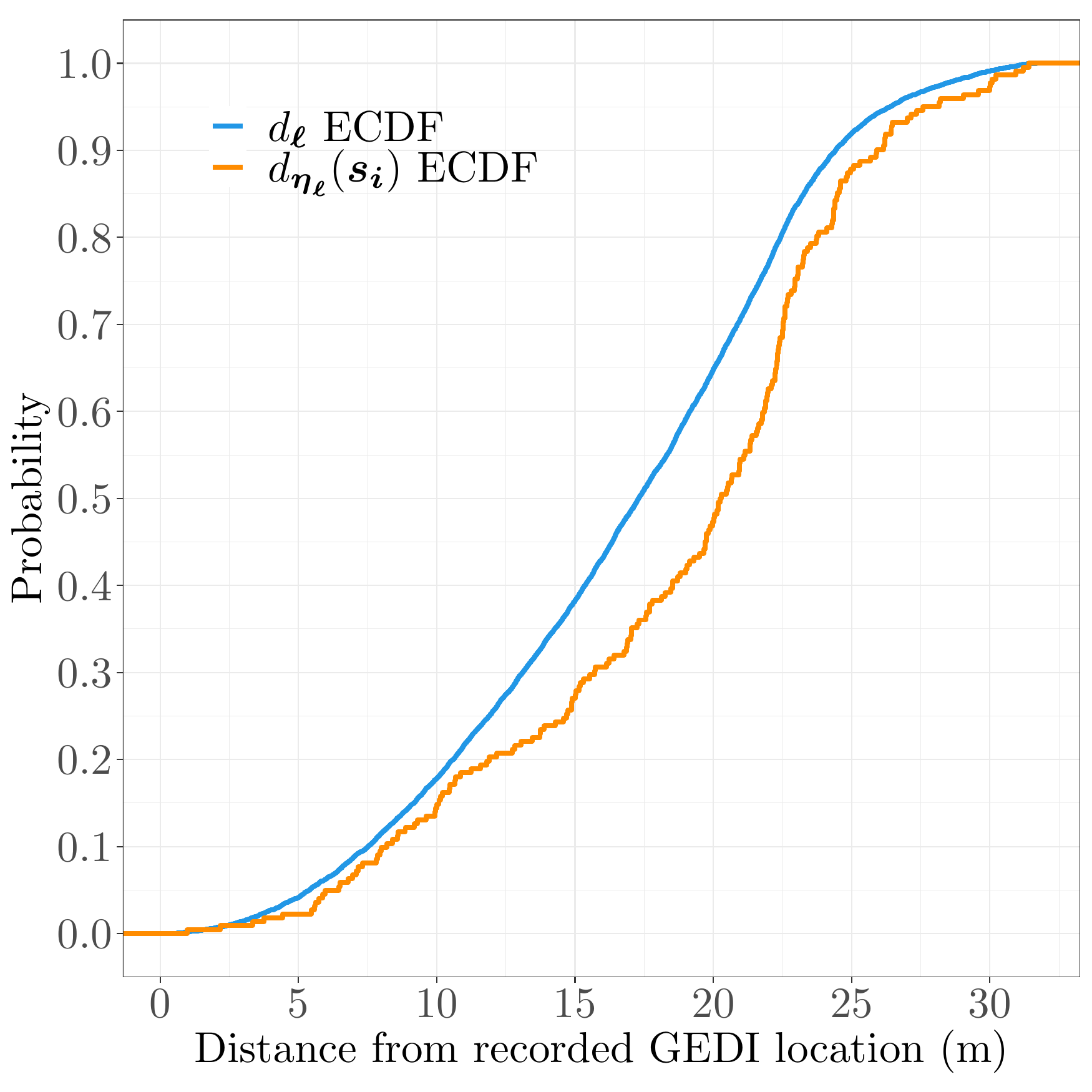} \label{fig:dist_ecdf}}
\caption{\subref{fig:dist_hist} Posterior distribution of $d_{\bell}$. \subref{fig:dist_ecdf} Empirical cumulative distribution function (ECDF) of $d_{\bell}$'s posterior distribution shown in \subref{fig:dist_hist} and the $n$ $d_{\bseta_{\bell}}(\bs_i)$.  }\label{fig:dists}
\end{center}
\end{figure}

Figure~\ref{fig:dist_hist} summarizes $d_{\bell}$'s estimated posterior distribution. This distribution, and its associated empirical cumulative distribution function (ECDF) given in Figure~\ref{fig:dist_ecdf}, suggest a relatively small portion of $d_{\bell}$'s posterior distribution is within 10 m of the reported footprint location. For example, the ECDF shows a $\sim$0.2 probability the true GEDI footprint center is within 10 m, $\sim$0.65 probability the true center is within 20 m, and $\sim$0.9 probability the true center is within 25 m.

While $d_{\bell}$'s distribution suggests potentially larger and more variable geolocation error beyond the estimated systematic error, it is important to recognize a limiting factor that prevents $\bell$ and hence $d_{\bell}$ from being a perfect summary of actual geolocation error. The issue arises when the simulator output at $\bs_i$ is the same, or very similar, to the output elsewhere in the focal area, and these outputs yield the largest log of the density (\ref{full_ltd}) values. In such settings, $\bell(\bs_i)$'s posterior distribution will have a central tendency over $\bs_i$ that might indicate an accurate GEDI location recording; however, the posterior will also have concentrations elsewhere in the focal area that yield similar $\bell(\bs_i)$ posterior density values. The result is a $d_{\bell}(\bs_i)$ concentration at zero and the other equally likely locations within the focal area. When this occurs, results are inconclusive (i.e., we cannot tell if there is geolocation error or not) and summaries such as given in Figure~\ref{fig:dists} include this ambiguity. 

\begin{figure}[!ht]
\begin{center}
\includegraphics[width=14cm,trim={0cm 0cm 0cm 0cm},clip]{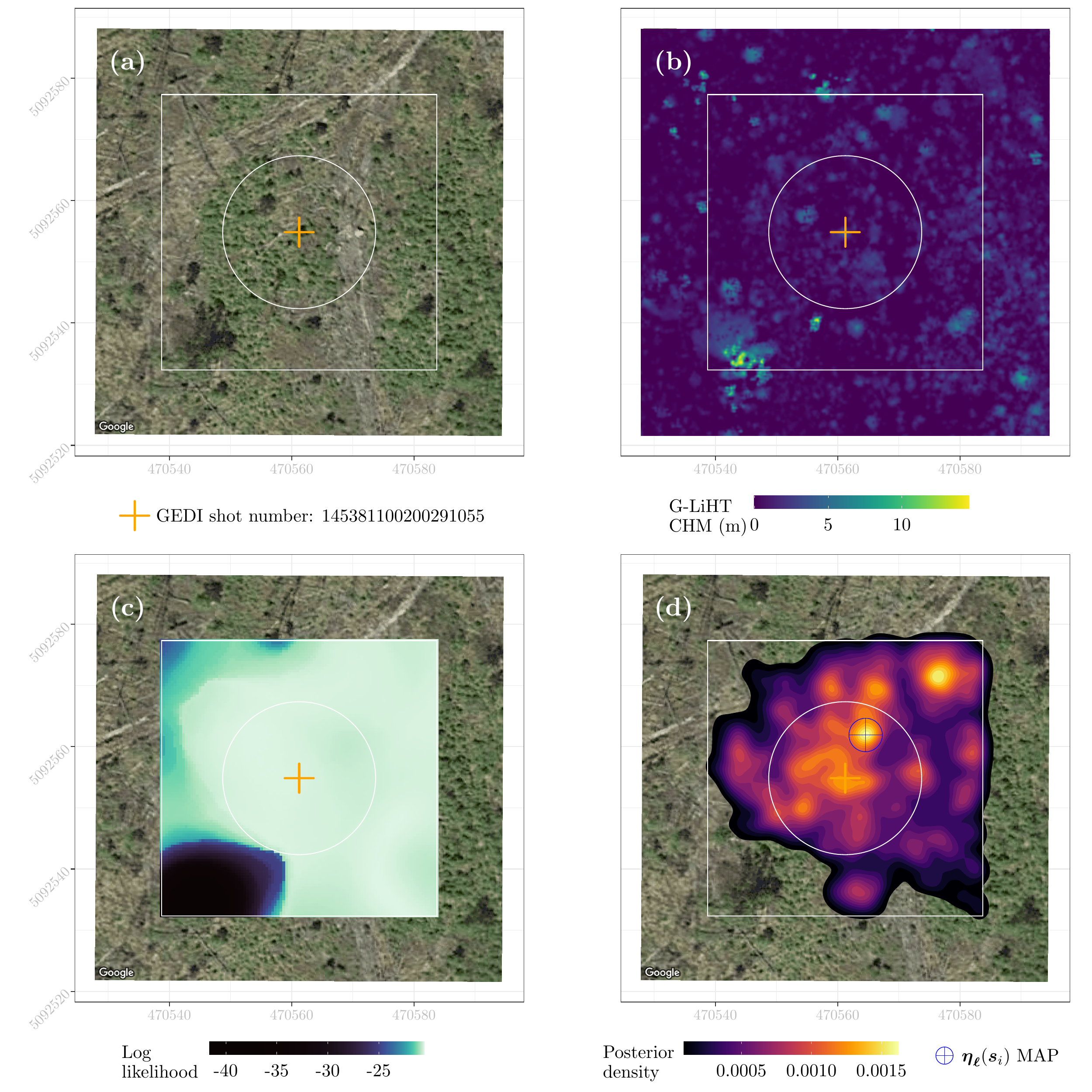}
\end{center}
\caption{Example focal area $i$=49 with (a) Google Satellite Imagery, (b) canopy height model derived from G-LiHT, (c) log likelihood surface calculated using parameter point estimates from the full model and (d) posterior distribution surface of $\bell(\bs_i)$ with MAP estimate of $\bseta_{\bell}(\bs_i)$. This focal area illustrates a structurally homogeneous area that results in a broad posterior distribution of $\bell(\bs_i)$.}\label{fig:fa_49}
\end{figure}

\begin{figure}[!ht]
\begin{center}
\includegraphics[width=14cm,trim={0cm 0cm 0cm 0cm},clip]{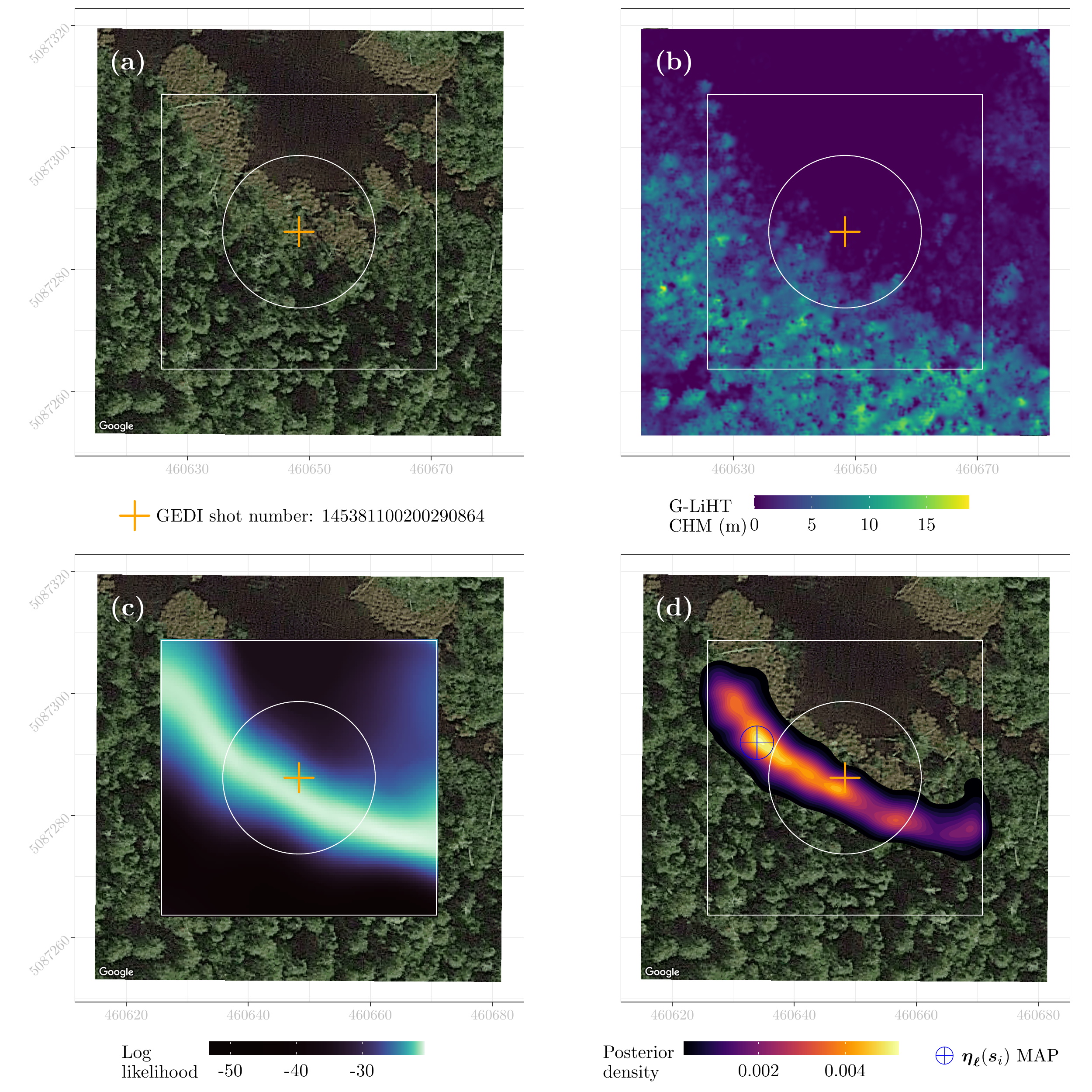}
\end{center}
\caption{Example focal area $i$=104 with (a) Google Satellite Imagery, (b) canopy height model derived from G-LiHT, (c) log likelihood surface calculated using parameter point estimates from the full model and (d) posterior distribution surface of $\bell(\bs_i)$ with MAP estimate of $\bseta_{\bell}(\bs_i)$. This focal area illustrates a structurally homogeneous edge that results in a long and narrow posterior distribution of $\bell(\bs_i)$.}\label{fig:fa_104}
\end{figure}

\begin{figure}[!ht]
\begin{center}
\includegraphics[width=7cm,trim={0cm 0cm 0cm 0cm},clip]{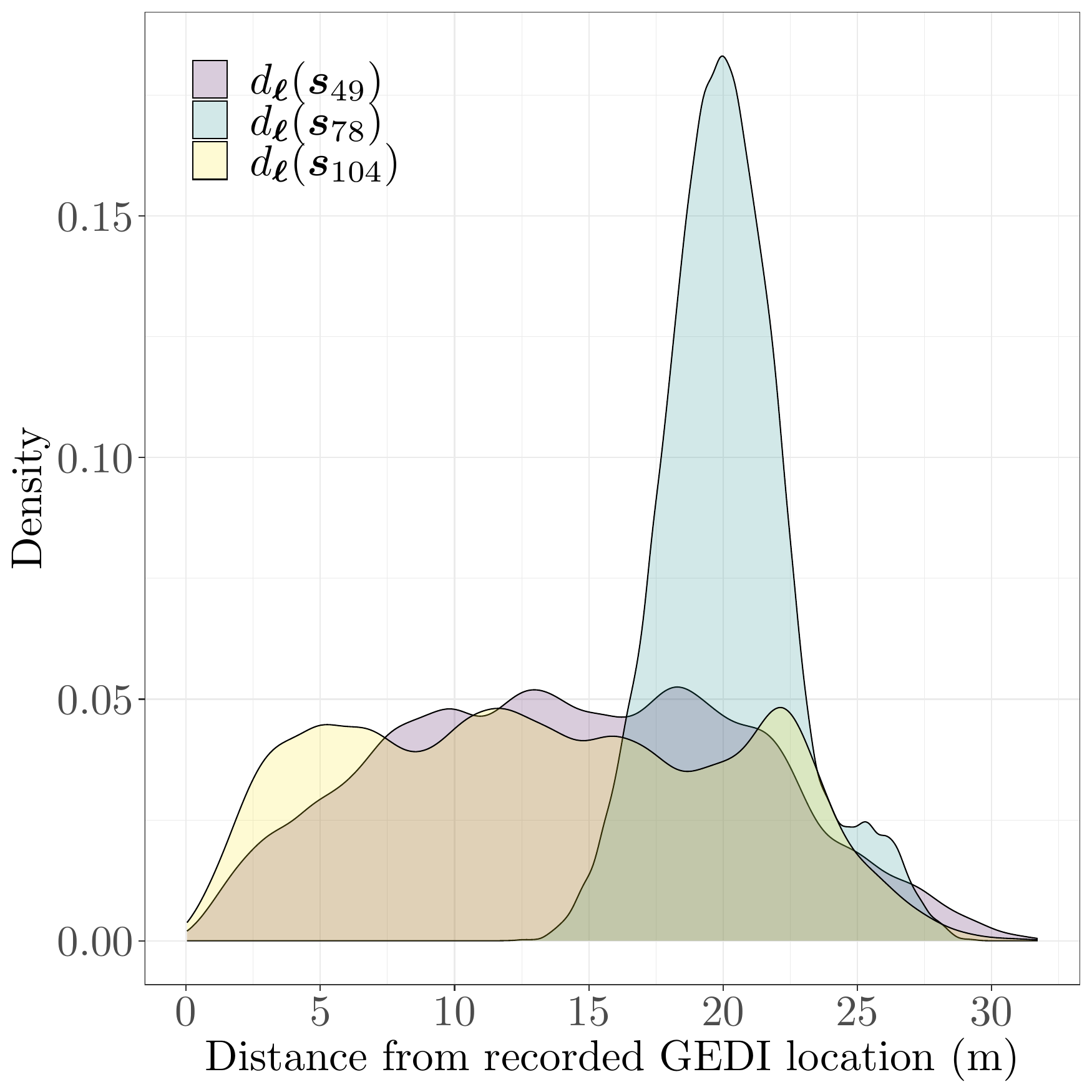}
\end{center}
\caption{Estimated posterior distributions of $d_{\bell}(\bs_i)$ for focal areas $i=(49,78,104)$ shown in Figures~\ref{fig:fa_49}, \ref{fig:fa_104}, and \ref{fig:fa_78}, respectively.}\label{fig:example_d_distributions}
\end{figure}

In our current analysis, the ambiguity described above occurs most often in focal areas with highly homogeneous vertical and horizontal vegetation structure. Such focal areas were noted by \cite{roy_et_al_2021} as forest types where GEDI geolocation error has smaller impact on the given analysis (which is an accurate assessment). For us, however, such settings result in inconclusive assessment of absolute geolocation error. Figures~\ref{fig:fa_49} and \ref{fig:fa_104} illustrate focal areas $i$=49 and $i$=104, respectively, which each exhibit homogeneous forest structure over $A_i$ and hence produce broad $d_{\bell}(\bs_i)$ posterior distributions as shown in Figure~\ref{fig:example_d_distributions}. 

A more appropriate measure of absolute geolocation error is the focal area specific MAP $\bseta_{\bell}(\bs_i)$ and associated $d_{\bseta_{\bell}}(\bs_i)$. These parameters do include the ambiguity associated with $d_{\bell}(\bs_i)$ (e.g., as illustrated in Figures~\ref{fig:fa_49}d and \ref{fig:fa_104}d), they simply capture the $\bell(\bs_i)$'s MAP and distance from this MAP to the recorded GEDI footprint center. Figure~\ref{fig:dist_ecdf} shows the ECDF for the $n$ $d_{\bseta_{\bell}}(\bs_i)$ estimates. This ECDF suggests a slightly larger absolute geolocation error than captured using $d_{\bell}$. For example, the $d_{\bseta_{\bell}}(\bs_i)$ based ECDF shows a $\sim$0.15 probability the true GEDI footprint center is within 10 m, $\sim$0.45 probability the true footprint center is within 20 m, and $\sim$0.9 probability the true footprint center is within 26 m.

\begin{figure}[!ht]
\begin{center}
\includegraphics[width=14cm,trim={0cm 0cm 0cm 0cm},clip]{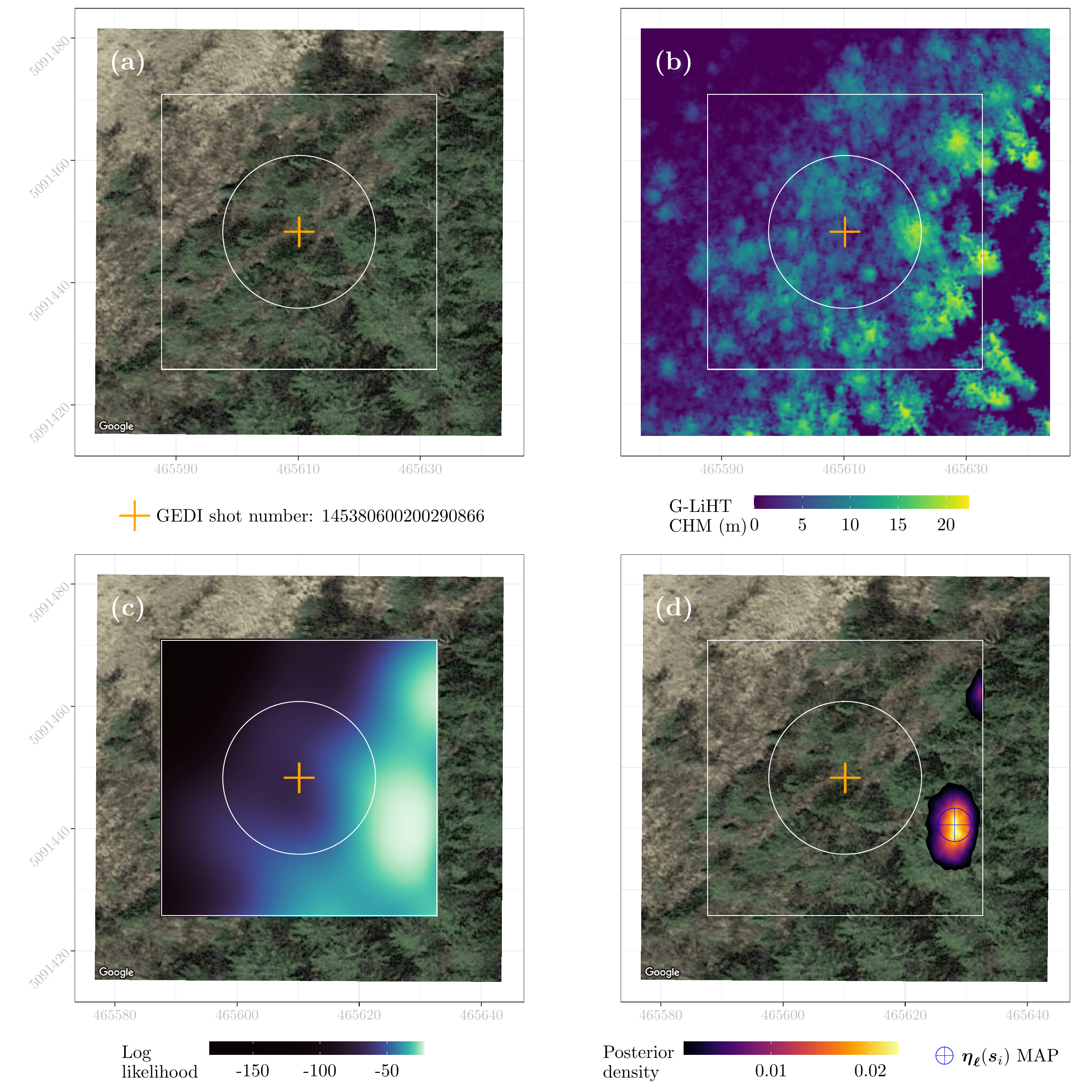}
\end{center}
\caption{Example focal area $i$=78. Illustration of posterior distribution of $\bell(\bs_i)$ concentrated well away from the reported footprint center location.}\label{fig:fa_78}
\end{figure}

As summarized in Figure~\ref{fig:dist_ecdf}, estimates of the ``true'' location from which $\bz(\bs_i)$ was measured are often relatively far from the recorded location. Examining each focal area's estimates (all $n$ figures of which are provided in Supplementary Material Figures~S2-S223) shows diverse posterior distributions across the $n$ $\bell(\bs_i)$. Some are quite broad, e.g., Figure~\ref{fig:fa_49}d, others follow contours in vertical forest structure which might or might not be apparent in the Google satellite imagery or G-LiHT data, e.g., Figure~\ref{fig:fa_104}, while others identify very specific locations that match well the observed $\bz(\bs_i)$, e.g., Figure~\ref{fig:fa_78}. In cases like focal area $i$=78, shown in Figure~\ref{fig:fa_78}, we are provided with a relatively precise geolocation error estimate, as reflected in the corresponding posterior distribution shown in Figure~\ref{fig:example_d_distributions}.

\section{Discussion}\label{sec:discussion}

All results are conditional on the posited model. By design the submodel can only estimate systematic geolocation error. As noted in Section~\ref{sec:submodel}, the submodel is a model-based analog to methods used in several similar studies that sought to estimate an ``optimal'' GEDI footprint location correction. These alternative methods simulated GEDI using ALS data for a finite number of locations at and around the recorded GEDI footprint center. The optimal location was then selected via an objective function that considered both the observed and simulated GEDI data. In spirit, our proposed method is similar to these previous approaches. Here, for the submodel, we discover the ``optimal'' location by maximizing the log of (\ref{sub_ltd}) and selecting a summary measure, e.g., mean, median, or MAP, of the estimated $\bell^\ast$'s posterior distribution. For the full model, maximizing the log of (\ref{full_ltd}) yields estimated posterior distributions of $\bell(\bs_i)$ for $i = (1, 2, \ldots, n)$ from which we can glean ``optimal'' systematic error correction via $\bseta_{\bell}$ and focal area specific error correction via $\bseta_{\bell}(\bs_i)$. By using a Bayesian inferential paradigm, we have access to full posterior distributions and can hence deliver full probabilistic error quantification for the $n$ $\bell(\bs_i)$, derived parameters (e.g., $d_{\bell}(\bs_i)$ and $\bz$ fitted values), and all other model parameters.

Our results suggest that the non-normality of $\bell$'s posterior distribution reduces the usefulness of estimating $\bmu_{\bell}$ in the full model, and that $\bseta_{\bell}$ is a more appropriate measure of systematic geolocation error. It is still useful to center the $\bell(\bs_i)$ on $\bs_i$ (which places prior weight on the assumption that the reported GEDI footprint center location is accurate). Hence, an attractive alternative specification of (\ref{l_model}) is $\bell(\bs_i) = \bs_i + \bepsilon_{\bell}(\bs_i)$ which replaces the last line in (\ref{full_ltd}) with $\prod^n_{i=1}\prod^2_{k=1}N\left(\ell_k(\bs_i) \given s_{i,k}, 1000\right)$. Removing $\bmu_{\bell}$ and associated hyperpriors yields a more parsimonious model and might, in some settings, improve numerical algorithm convergence used to estimate parameters. We have fit this alternative specification and it yields nearly identical results to the full model detailed in Section~\ref{sec:models}.

There is geolocation error in the reported GEDI data considered here and summaries of systematic geolocation error do not reflect well the actual error apparent in most focal areas (i.e., those focal areas where the $\bell(\bs_i)$'s posterior measure of central tendency is distinct from $\bs_i$). Attempting to address geolocation error by shifting GEDI footprint locations to the $\bell^\ast$'s mean, median, or MAP will provide only marginal improvement in fit as reflected by the middle subplot in Figure~\ref{fig:obs_vs_fitted}. While there was clear systematic error, there was also large variation in actual error across GEDI footprint locations. This large variation has implications for studies that attempt to spatially coincide GEDI data with other remotely sensed or georeferenced field measurements.

\section{Summary and future work}\label{sec:conclusion}

We propose a general model to estimate geolocation error in spaceborne sampling LiDAR using a simulator function that takes a set of potentially unknown parameters and spatially coinciding high spatial accuracy ALS data. The model accommodates multiple canopy height metrics, can estimate unknown parameters in the simulator function, and provides additive and multiplicative correction when mapping between the observed and simulated LiDAR metrics. The key feature of the proposed model is its ability to deliver probabilistic estimates of actual and systematic geolocation error via a rich, flexible, and extendable Bayesian framework. 

We motivate and illustrate the proposed model using GEDI and G-LiHT ALS data from Northern Maine, USA. Our results indicate some degree of systematic geolocation error over all GEDI footprint locations, with more variability in the actual geolocation error across locations. For this initial study, the spatial and temporal scope were kept intentionally narrow to minimize confounding effects of spatial and temporal variability. In the future, the full or submodel can be extended to include both spatial or temporal random effects on $\bell(\bs_i)$ and $\bell^\ast$, respectively, designed to capture differences among acquisitions across locations and dates. The full model random effects can also be written to accommodate spatially and/or temporally structured $\bell(\bs_i)$ that would allow for estimation of correlation among actual errors. If such spatial or temporal correlation exists, then the full model could be used to correct geolocation error in regions where no ALS data exist.

\section*{Acknowledgements}
Funding was provided by: NASA Carbon Monitoring System (CMS) grants Hayes (CMS 2020) and Cook (CMS 2018), National Science Foundation (NSF) DMS-1916395, and a joint venture agreement with the USDA Forest Service Forest Inventory and Analysis.

\newpage

\begin{thebibliography}{}

\bibitem[Blair and Hoften, 1999]{blair_and_hoften_1999}
Blair, J.~B. and Hoften, M.~A. (1999).
\newblock Modeling laser altimeter return waveforms over complex vegetation
  using high-resolution elevation data.
\newblock {\em Geophysical Research Letters}, 26(16):2509--2512.

\bibitem[Cohen et~al., 2020]{cohen_et_al_2020}
Cohen, W., Yang, Z., Healey, S., and Andersen, H. (2020).
\newblock Disturbance history and forest biomass from landsat for six us sites,
  1985-2014. {ORNL} {D}istributed {A}ctive {A}rchive {C}enter. doi:
  10.3334/ornldaac/1679.

\bibitem[Cook et~al., 2013]{cook_et_al_2013}
Cook, B.~D., Corp, L.~A., Nelson, R.~F., Middleton, E.~M., Morton, D.~C.,
  McCorkel, J.~T., Masek, J.~G., Ranson, K.~J., Ly, V., and Montesano, P.~M.
  (2013).
\newblock {NASA} {G}oddard’s {L}idar, {H}yperspectral and {T}hermal
  ({G-LiHT}) {A}irborne {I}mager.
\newblock {\em Remote Sensing}, 5(8):4045--4066.

\bibitem[Deo et~al., 2017]{deo_et_al_2017}
Deo, R.~K., Russell, M.~B., Domke, G.~M., Andersen, H.-E., Cohen, W.~B., and
  Woodall, C.~W. (2017).
\newblock Evaluating site-specific and generic spatial models of aboveground
  forest biomass based on landsat time-series and lidar strip samples in the
  eastern usa.
\newblock {\em Remote Sensing}, 9(6).

\bibitem[Dubayah et~al., 2020]{dubayah_et_al_2020}
Dubayah, R., Blair, J.~B., Goetz, S., Fatoyinbo, L., Hansen, M., Healey, S.,
  Hofton, M., Hurtt, G., Kellner, J., Luthcke, S., Armston, J., Tang, H.,
  Duncanson, L., Hancock, S., Jantz, P., Marselis, S., Patterson, P.~L., Qi,
  W., and Silva, C. (2020).
\newblock The global ecosystem dynamics investigation: High-resolution laser
  ranging of the earth’s forests and topography.
\newblock {\em Science of Remote Sensing}, 1:100002.

\bibitem[Dubayah et~al., 2021]{dubayah_et_al_2021_gedi_dataset}
Dubayah, R., Hofton, M., Blair, J., Armston, J., Tang, H., and Luthke, S.
  (2021).
\newblock {GEDI L2A} elevation and height metrics data global footprint level
  v002 [data set].
\newblock NASA EOSDIS Land Processes DAAC.

\bibitem[Duncanson et~al., 2022]{duncanson_et_al_2022}
Duncanson, L., Kellner, J.~R., Armston, J., Dubayah, R., Minor, D.~M., Hancock,
  S., Healey, S.~P., Patterson, P.~L., Saarela, S., Marselis, S., Silva, C.~E.,
  Bruening, J., Goetz, S.~J., Tang, H., Hofton, M., Blair, B., Luthcke, S.,
  Fatoyinbo, L., Abernethy, K., Alonso, A., Andersen, H.-E., Aplin, P., Baker,
  T.~R., Barbier, N., Bastin, J.~F., Biber, P., Boeckx, P., Bogaert, J.,
  Boschetti, L., Boucher, P.~B., Boyd, D.~S., Burslem, D.~F., Calvo-Rodriguez,
  S., Chave, J., Chazdon, R.~L., Clark, D.~B., Clark, D.~A., Cohen, W.~B.,
  Coomes, D.~A., Corona, P., Cushman, K., Cutler, M.~E., Dalling, J.~W.,
  Dalponte, M., Dash, J., de~Miguel, S., Deng, S., Ellis, P.~W., Erasmus, B.,
  Fekety, P.~A., Fernandez-Landa, A., Ferraz, A., Fischer, R., Fisher, A.~G.,
  García-Abril, A., Gobakken, T., Hacker, J.~M., Heurich, M., Hill, R.~A.,
  Hopkinson, C., Huang, H., Hubbell, S.~P., Hudak, A.~T., Huth, A., Imbach, B.,
  Jeffery, K.~J., Katoh, M., Kearsley, E., Kenfack, D., Kljun, N., Knapp, N.,
  Král, K., Krůček, M., Labrière, N., Lewis, S.~L., Longo, M., Lucas,
  R.~M., Main, R., Manzanera, J.~A., Martínez, R.~V., Mathieu, R., Memiaghe,
  H., Meyer, V., Mendoza, A.~M., Monerris, A., Montesano, P., Morsdorf, F.,
  Næsset, E., Naidoo, L., Nilus, R., O’Brien, M., Orwig, D.~A.,
  Papathanassiou, K., Parker, G., Philipson, C., Phillips, O.~L., Pisek, J.,
  Poulsen, J.~R., Pretzsch, H., Rüdiger, C., Saatchi, S., Sanchez-Azofeifa,
  A., Sanchez-Lopez, N., Scholes, R., Silva, C.~A., Simard, M., Skidmore, A.,
  Stereńczak, K., Tanase, M., Torresan, C., Valbuena, R., Verbeeck, H., Vrska,
  T., Wessels, K., White, J.~C., White, L.~J., Zahabu, E., and Zgraggen, C.
  (2022).
\newblock Aboveground biomass density models for {NASA}’s {G}lobal
  {E}cosystem {D}ynamics {I}nvestigation ({GEDI}) lidar mission.
\newblock {\em Remote Sensing of Environment}, 270:112845.

\bibitem[Frazer et~al., 2011]{frazer_et_al_2011}
Frazer, G., Magnussen, S., Wulder, M., and Niemann, K. (2011).
\newblock Simulated impact of sample plot size and co-registration error on the
  accuracy and uncertainty of lidar-derived estimates of forest stand biomass.
\newblock {\em Remote Sensing of Environment}, 115(2):636--649.

\bibitem[Gelman et~al., 2013]{gelman2013}
Gelman, A., Carlin, J., Stern, H., Dunson, D., Vehtari, A., and Rubin, D.
  (2013).
\newblock {\em Bayesian Data Analysis, Third Edition}.
\newblock Chapman \& Hall/CRC Texts in Statistical Science. Taylor \& Francis.

\bibitem[Goward et~al., 2008]{goward_et_al_2008}
Goward, S.~N., Masek, J.~G., Cohen, W., Moisen, G., Collatz, G.~J., Healey, S.,
  Houghton, R.~A., Huang, C., Kennedy, R., Law, B., Powell, S., Turner, D., and
  Wulder, M.~A. (2008).
\newblock Forest disturbance and north american carbon flux.

\bibitem[Hancock et~al., 2019]{hancock_et_al_2019}
Hancock, S., Armston, J., Hofton, M., Sun, X., Tang, H., Duncanson, L.~I.,
  Kellner, J.~R., and Dubayah, R. (2019).
\newblock The {GEDI} simulator: A large-footprint waveform lidar simulator for
  calibration and validation of spaceborne missions.
\newblock {\em Earth and Space Science}, 6(2):294--310.

\bibitem[Lang et~al., 2022]{lang_et_al_2022}
Lang, N., Kalischek, N., Armston, J., Schindler, K., Dubayah, R., and Wegner,
  J.~D. (2022).
\newblock Global canopy height regression and uncertainty estimation from
  {GEDI} lidar waveforms with deep ensembles.
\newblock {\em Remote Sensing of Environment}, 268:112760.

\bibitem[Liu et~al., 2021]{liu_et_al_2021}
Liu, A., Cheng, X., and Chen, Z. (2021).
\newblock Performance evaluation of {GEDI} and {ICESat-2} laser altimeter data
  for terrain and canopy height retrievals.
\newblock {\em Remote Sensing of Environment}, 264:112571.

\bibitem[Milenkovi\'{c} et~al., 2017]{milenkovic_et_al_2017}
Milenkovi\'{c}, M., Schnell, S., Holmgren, J., Ressl, C., Lindberg, E.,
  Hollaus, M., Pfeifer, N., and Olsson, H. (2017).
\newblock Influence of footprint size and geolocation error on the precision of
  forest biomass estimates from space-borne waveform lidar.
\newblock {\em Remote Sensing of Environment}, 200:74--88.

\bibitem[Quirós et~al., 2021]{quiros_et_al_2021}
Quirós, E., Polo, M.-E., and Fragoso-Campón, L. (2021).
\newblock {GEDI} elevation accuracy assessment: A case study of southwest
  spain.
\newblock {\em IEEE Journal of Selected Topics in Applied Earth Observations
  and Remote Sensing}, 14:5285--5299.

\bibitem[Roy et~al., 2021]{roy_et_al_2021}
Roy, D., Kashongwe, H., and Armston, J. (2021).
\newblock The impact of geolocation uncertainty on {GEDI} tropical forest
  canopy height estimation and change monitoring.
\newblock {\em Science of Remote Sensing}, 4(100024).

\bibitem[Saarela et~al., 2018]{saarela_et_al_2018}
Saarela, S., Holm, S., Healey, S.~P., Andersen, H.-E., Petersson, H., Prentius,
  W., Patterson, P.~L., Næsset, E., Gregoire, T.~G., and Ståhl, G. (2018).
\newblock Generalized hierarchical model-based estimation for aboveground
  biomass assessment using {GEDI} and {L}andsat data.
\newblock {\em Remote Sensing}, 10(11).

\bibitem[Silva et~al., 2021]{silva_et_al_2021}
Silva, C.~A., Duncanson, L., Hancock, S., Neuenschwander, A., Thomas, N.,
  Hofton, M., Fatoyinbo, L., Simard, M., Marshak, C.~Z., Armston, J., Lutchke,
  S., and Dubayah, R. (2021).
\newblock Fusing simulated {GEDI}, {ICESat-2} and {NISAR} data for regional
  aboveground biomass mapping.
\newblock {\em Remote Sensing of Environment}, 253:112234.

\bibitem[Tak et~al., 2018]{tak_et_al_2018}
Tak, H., Meng, X.-L., and van Dyk, D.~A. (2018).
\newblock A repelling–attracting metropolis algorithm for multimodality.
\newblock {\em Journal of Computational and Graphical Statistics},
  27(3):479--490.

\bibitem[Venables and Ripley, 2002]{venables_ripley_2002}
Venables, W.~N. and Ripley, B.~D. (2002).
\newblock {\em Modern Applied Statistics with S}.
\newblock Springer, New York, fourth edition.
\newblock ISBN 0-387-95457-0.

\bibitem[Wang et~al., 2022]{wang_et_al_2021}
Wang, C., Elmore, A.~J., Numata, I., Cochrane, M.~A., Shaogang, L., Huang, J.,
  Zhao, Y., and Li, Y. (2022).
\newblock Factors affecting relative height and ground elevation estimations of
  {GEDI} among forest types across the conterminous usa.
\newblock {\em GIScience \& Remote Sensing}, 59(1):975--999.

\bibitem[Zhang, 2016]{zhang13}
Zhang, X. (2016).
\newblock An optimized blas library based on gotoblas2.
\newblock \url{https://github.com/xianyi/OpenBLAS/}.
\newblock Accessed 2015-06-01.

\end{thebibliography}

\end{document}